\documentclass[11pt]{article}
\usepackage{float}
\usepackage{authblk}
\usepackage{amssymb,amsmath}
\usepackage{graphicx}
\usepackage{caption}
\usepackage{subcaption}
\usepackage{cite}
\usepackage{color}
\usepackage{setspace}
\usepackage[colorlinks=true
,urlcolor=blue
,citecolor=blue
,linkcolor=blue
,pagecolor=blue
,linktocpage=true
,pdfproducer=medialab
]{hyperref}
\usepackage[a4paper,width=17cm,height=25cm]{geometry}
\usepackage{morefloats}
\usepackage{siunitx}

\makeatletter \renewcommand{\@dotsep}{10000} \makeatother
\title{Testing Leptoquark and $Z^{\prime}$ Models via $B\to K_{1}(1270,1400)\mu^{+}\mu^{-}$ Decays}
\author[a] {Zhuo-Ran Huang%
	\thanks{\texttt huangzr@ihep.ac.cn }}
\author[b,a] {Muhammad Ali Paracha%
	\thanks{\texttt aliparacha@sns.nust.edu.pk}}
\author[c]{Ishtiaq Ahmed%
	\thanks{\texttt ishtiaqmusab@gmail.com}}
\author[a,d] {Cai-Dian L\"u%
	\thanks{\texttt lucd@ihep.ac.cn}}
	
\affil[a]{Institute of High Energy Physics, Chinese Academy of Sciences, Beijing 100049, China}
\affil[b]{Department of Physics, School of Natural Sciences (SNS), National University of Sciences and Technology (NUST), Sector H-12 Islamabad, Pakistan.}
\affil[c]{National Centre for Physics, Islamabad, Pakistan.}
\affil[d]{School of Physics, University of Chinese Academy of Sciences, Beijing 100049, China}
\begin{document}
\maketitle
\begin{abstract}
The measurements of $R_{K^{(*)}}=\mathcal B(B\to K^{(*)}\mu^{+}\mu^{-})/\mathcal B(B\to K^{(*)}e^{+}e^{-})$ in recent years have hinted lepton flavor non-universality and thus drawn widespread attentions. If these anomalies are induced by new physics (NP), deviations from the SM predictions may also be found in other channels via the same process at the quark level. In this work, we study in $B\to K_{1}(1270,1400)\mu^{+}\mu^{-}$ decays the effects of two popular classes of NP models which can address the $b\to s$ anomalies, i.e. the leptoquark models and the $Z^{\prime}$ models. By assuming that NP only affects the $b\to s\mu^+\mu^-$ transition, we find that the unpolarized and polarized lepton flavor universality (LFU) ratios $R_{K_1^{(L,T)}(1270)}$ are useful to distinguish among the NP models (scenarios) and the SM because they are sensitive to the NP effects and insensitive to the mixing angle $\theta_{K_1}$, while the $R_{K_1^{(L,T)}(1400)}$ are sensitive to both NP and $\theta_{K_1}$. Another ratio $R_\mu(K_1)=\mathcal B(B\to K_1(1400)\mu^+\mu^-)/\mathcal B(B\to K_1(1270)\mu^+\mu^-)$ is shown to depend weekly on the effects from the NP models (scenarios) under consideration, and thus can be used to determine the $\theta_{K_1}$ and complement the $R_{K_1^{(L,T)}}$ in the probe for NP.
\end{abstract}

\newpage
\section{Introduction}\label{intro}
In the past few years, several anomalies in B physics \cite{Li:2018lxi,Bifani:2018zmi} have been heatedly discussed in the high-energy physics community since these measurements are hints of new physics beyond the Standard Model (SM) or more precisely, the lepton flavor universality violation (LFUV). The BaBar Collaboration~\cite{BAB,BB1} first reported one class of such anomalies, in the measurement of
\begin{eqnarray}
 R_{D^{(\ast)}}=\frac{\mathcal B(B\to D^{(\ast)}\tau\bar\nu)}{\mathcal B(B\to D^{(\ast)}\ell\bar\nu)}\,.\label{RD1}
\end{eqnarray}
The main advantage of considering such a ratio is that it cancels exactly the Cabibbo-Kobayashi-Maskawa matrix (CKM) elements and the uncertainties due to the transition form factors are also partially but largely cancelled.
Later on Belle~\cite{Huschle:2015rga,Sato:2016svk,Hirose:2016wfn} and LHCb~\cite{Aaij:2015yra,Aaij:2017uff,Aaij:2017deq,Aaij:2017tyk} measured the same ratio and also observed the excess: the measured
value of $R^{\text{exp}}_{D^{\ast}}$ is greater than $R^{\text{SM}}_{D^{(\ast)}}$ prediction. The most recent values of $ R_{D^{(\ast)}}$ given by the Heavy Flavor Averaging Group (HFLAV)~\cite{HFLAV} are
\begin{eqnarray}
 R_{D}=0.407(46)\,,~~~R_{D^{\ast}}=0.306(15)\,,
\end{eqnarray}
The difference between the SM predictions~\cite{Aebischer:2018iyb,Huang:2018nnq,Fajfer:2012vx,Iguro:2018vqb,Li:2016vvp,Sakaki:2013bfa,Jaiswal:2017rve} and experimental values is approximately $3-4\sigma$ and thus gives a hint of NP.

Apart from the tree-level charged current semileptonic B decays, the loop-level rare $B$ decays mediated by the flavor-changing neutral current (FCNC) transition $b\to s\ell^{+}\ell^{-}$ also give hints of lepton flavor non-universality. Such measurements include the LFU ratios
\begin{equation}
 R_{K^{(\ast)}}=\frac{\int_{q^{2}_{min}}^{q^{2}_{max}}\frac{d\mathcal B(B\to K^{(\ast)}\mu^{+}\mu^{-})}{dq^{2}}dq^{2}}{\int_{q^{2}_{min}}^{q^{2}_{max}}\frac{d\mathcal B(B\to K^{(\ast)}e^{+}e^{-})}{dq^{2}}dq^{2}}\,,
\end{equation}
and the values reported by LHCb in different bins are \cite{Aaij:2014ora,Aaij:2017vbb}
\begin{eqnarray}
 R_{K}^{[1,6]} = 0.745 ^{+0.090}_{-0.074} \pm 0.036\,,R_{K^*}^{[0.045, 1.1]} = 0.66 ^{+0.11}_{-0.07} \pm 0.03\,,R_{K^*}^{[1.1, 6]} = 0.69 ^{+0.11}_{-0.07} \pm 0.05\,,
\end{eqnarray}
of which the tensions with the SM predictions are respectively $2.6\sigma$, $2.1-2.3\sigma$ and $2.6\sigma$ \cite{1811.11309}. These ratios have theoretical uncertainties that are almost canceled (less than $1\%$ \cite{Bordone:2016gaq}), making them very clean probe for NP/LFUV \cite{Hiller:2003js}. Although in principal NP is possible to affect both $b\to s e^+e^-$ and $b\to s \mu^+\mu^-$, in many existing studies \cite{Altmannshofer:2017yso,Alok:2017jgr,DL,Capdevila:2017bsm,Kumar:2018kmr,Hiller:2017bzc,Sala:2017ihs,Ciuchini:2017mik} the assumption that NP only affects $b\to s \mu^+\mu^-$ has been considered because several other deviations from the SM in $b \to s\mu^+\mu^-$ have been observed \cite{Altmannshofer:2017fio,Alok:2010zd,Alok:2011gv} and the measured branching fraction of $B\to Ke^{+}e^{-}$ is consistent with the SM prediction. 

Among the NP models that can explain the $b\to s\mu^+\mu^-$ data are leptoquark models \cite{CCO,AGC,HS1,GNR,VH,SM,FK,BFK,BKSZ} and  $Z^{\prime}$ models \cite{Altmannshofer:2014cfa,Crivellin:2015mga,Altmannshofer:2015mqa,Crivellin:2016ejn,Ahmed:2017vsr,Chala:2018igk}. In the language of the effective field theory, these NP models can modify the Wilson coefficients so that the effective Hamiltonian fulfills one of the three possible model-independent NP scenarios that can explain the $b\to s\mu^+\mu^-$ data \cite{Capdevila:2017bsm}. If these NP models or model-independent explanations depict the NP in $b\to s\mu^+\mu^-$ at the quark level, one naturally expects to observe similar anomalies in other rare decays such as  $B\to K_1\mu^+\mu^-$.\footnote{For studies of the lepton flavor universality in various $b\to s\mu^+\mu^-$ channels, see \cite{Dutta:2019wxo,Wang:2017mrd}.}
In this work, we extent the study of the NP in $b\to s\mu^+\mu^-$ to axial-vector final state mesons, i.e. the $K_1$ states, which should be useful to test the existing model-independent and model explanations of the $b\to s\mu^+\mu^-$ anomalies. In this context $B\to K_{1}(1270,1400)\ell^{+}\ell^{-}$ decays are prosperous
in phenomenology \cite{Hatanaka:2008xj,Hatanaka:2008gu,Paracha:2007yx,Ahmed:2008ti,Ahmed:2010tt,Ahmed:2011vr,Saddi1,Saddi2,Li:2009rc,Li:2011nf} as the physical states $K_{1}(1270)$ and $K_1(1400)$ are mixture of $^3P_1$ and $^1P_1$ states $K_{1A}$ and $K_{1B}$:
\begin{eqnarray}
 |K_{1}(1270)\rangle&=&|K_{1A}\rangle\sin\theta_{K_{1}}+|K_{1B}\rangle\cos\theta_{K_{1}}\,,\label{theta}\\
 |K_{1}(1400)\rangle&=&|K_{1A}\rangle\cos\theta_{K_{1}}-|K_{1B}\rangle\sin\theta_{K_{1}}\,.\label{theta1}
\end{eqnarray}
The mixing angle $\theta_{K_1}$ has not been precisely determined, and its was estimated to be  $-(34\pm 13)^{\circ}$ from the decay $B\to K_{1}(1270)\gamma$ and $\tau\to K_{1}(1270)\nu_{\tau}$ \cite{Hatanaka:2008xj}. Therefore in this work we consider different possibilities for $\theta_{K_1}$.

It has been found in articles and also by our independent study that the observables like the branching ratio, the different polarization and angular observables and also the LFU ratios for semileptonic $B$ meson decays are greatly influenced in different NP models. However predictions for many of these observables can have large theoretical uncertainties, which makes it more involved to distinguish NP. Hence in this work we mainly concentrate on the LFU ratios $R_{K_1}$, for both unpolarized and polarized $K_1$ final states. Numerically we use the Wilson coefficients and the NP parameters in $Z^{'}$ and leptoquark models obtained from the fits to the $b\to s\mu^+\mu^-$ data (including the branching fractions and the angular observables for $B\to K^*\mu^+\mu^-$ and $B_s\to \phi^*\mu^+\mu^-$ as well as the $R_{K^{(*)}}$) in \cite{DL} to provide with predictions for the LFU ratios, which can be tested by future experiments to dig out the status of NP/LFUV. Since some of the obtained ratios are sensitive to $\theta_{K_1}$, as a complementary study of NP we also perform an analysis of the ratio $R_\mu(K_1)=\mathcal B(B\to K_1(1400)\mu^+\mu^-)/\mathcal B(B\to K_1(1270)\mu^+\mu^-)$ that has been found to be insensitive to the NP effects from a single NP operator \cite{Hatanaka:2008gu}, which could be useful to determine the mixing angle.

The organization of the paper is as follows. In Sec.~\ref{sec2} we present the theoretical formalism in the language of the effective field theory, including giving a brief review of the model-independent NP scenarios, the leptoquark models and the $Z^\prime$ models. In Sec.~\ref{sec3} and Sec.~\ref{sec4}, we respectively describe the hadronic form factors adopted in this work and the physical observables. In Sec.\ref{sec5}, we present our predictions for different unpolarized and polarized ratios. At last in Sec.~\ref{sec5} we give our summary and conclusions.
\section{Theoretical Tool Kit}\label{sec2}
In this section we briefly discuss the theoretical setup and new physics models to analyze the physical observables in
$B^{0}_{d}\to K_{1}(1270,1400)\mu^{+}\mu^{-}$ decays, more precisely we focus our attention on lepton flavor universality parameters for both polarized and unpolarized final state axial vector meson $K_{1}(1270,1400)$.

The basic ingredient to do phenomenology in rare decays is the effective Hamiltonian, which for the $b\to s\mu^{+}\mu^{-}$ process at the quark level can be written as
\begin{eqnarray}
 H_{eff}=-\frac{4 G_{F}}{\sqrt{2}}V_{tb}V^{\ast}_{ts}\left[\sum_{i=1}^{6}C_{i}(\mu)O_{i}(\mu)+\sum_{i=7,9,10}C_{i}(\mu)O_{i}(\mu)
+C_{i}^{\prime}(\mu)O_{i}^{\prime}(\mu) \right]\,.\label{H1}
\end{eqnarray}
The effective Hamiltonian given in Eq.(\ref{H1}) contains the four quark and electromagnetic operators $O_{i}$, and $C_{i}(\mu)$
are their corresponding Wilson coefficients. $G_{F}$ is the Fermi coupling constant, and $V_{tb}$ and $V_{ts}^{\ast}$ are the CKM matrix elements.

The effective operators that contributes both in SM and in NP are summarized as follows
\begin{eqnarray}
 O_{7} &=&\frac{e^{2}}{16\pi ^{2}}m_{b}\left( \bar{s}\sigma _{\mu \nu }P_{R}b\right) F^{\mu \nu }\,,  \notag \\
O_{9} &=&\frac{e^{2}}{16\pi ^{2}}(\bar{s}\gamma _{\mu }P_{L}b)(\bar{l}\gamma^{\mu }l)\,,  \label{op-form} \\
O_{10} &=&\frac{e^{2}}{16\pi ^{2}}(\bar{s}\gamma _{\mu }P_{L}b)(\bar{l} \gamma ^{\mu }\gamma _{5} l)\,,  \notag
\end{eqnarray}
The primed operators given in Eq.(\ref{H1}) are obtained by replacing left-handed fields ($L$) with right-handed ($R$) ones. In this work we only consider those
scenarios of NP where the operator basis remains the same as that of SM but the Wilson coefficients get modified. The modified Wilson coefficients in the above Hamiltonian can be written as
\begin{eqnarray}
 C_{9}^{\text{tot}}=C_{9}^{\text{eff}}+C_{9}^{\text{NP}}\label{WC1}\\
 C_{10}^{\text{tot}}=C_{10}^{\text{SM}}+C_{10}^{\text{NP}}\label{WC2}
\end{eqnarray}
The Wilson coefficients incoroporate the short distance physics and are evaluated through perturbative approach. The factorizable
contributions from current-current, QCD penguins and chromomagnetic operators $O_{1-6,8}$ have been consolidated in the Wilson coefficients $C_{9}^{\text{eff}}(s)$ and $C_{7}^{\text{eff}}(s)$
and their experssions are given as\cite{Du}
\begin{eqnarray}
 C_{7}^{\text{eff}}(q^{2})=C_{7}-\frac{1}{3}(C_{3}+\frac{4}{3}C_{4}+20C_{5}+\frac{80}{3}C_{6})-\frac{\alpha_{s}}{4\pi}[(C_{1}-6C_{2})F^{(7)}_{1,c}(q^{2})+C_{8}F^{7}_{8}(q^{2})]\notag\\
 C_{9}^{\text{eff}}(q^{2})=C_{9}+\frac{4}{3}(C_{3}+\frac{16}{3}C_{5}+\frac{16}{9}C_{6})-h(0,q^{2})(\frac{1}{2}C_{3}+\frac{2}{3}C_{4}+8C_{5}+\frac{32}{3}C_{6})\notag\\
-(\frac{7}{2}C_{3}+\frac{2}{3}C_{4}+38C_{5}+\frac{32}{3}C_{6})h(m_{b},q^{2})+(\frac{4}{3}C_{1}+C_{2}+6C_{3}+60C_{5})h(m_{c},q^{2})\notag\\
-\frac{\alpha_{s}}{4\pi}[C_{1}F^{(9)}_{1,c}(q^{2})+C_{2}F^{(9)}_{2,c}(q^{2})+C_{8}F^{(9)}_{8}(q^{2})]\label{WC3}
\end{eqnarray}
The Wilson coefficients given in Eq.(\ref{WC3}) involves the functions $h(m_{q},s)$ with $q=c,b$ and functions $F^{7,9}_{8}(q^{2})$ are
defined in\cite{BF} and the function $F^{(7,9}_{1,c}(q^{2})$ given in \cite{AH} for low $q^{2}$ and in \cite{GC} for high $q^{2}$.
The numerical of Wilson coefficents $C_{i}(i=1,.........,10)$ at $\mu\sim m_{b}$ are presented in Table-\ref{wc table}.
\begin{table*}[ht]
\begin{tabular}{cccccccccc}
\hline\hline
$C_{1}$&$C_{2}$&$C_{3}$&$C_{4}$&$C_{5}$&$C_{6}$&$C_{7}$&$C_{9}$&$C_{10}$
\\ \hline
  \ \ -0.263 \ \ &  \ \ 1.011 \ \ & \ \ 0.005 \ \ &  \ \ -0.0806  \ \ &  \ \ 0.0004 \ \ &  \ \ 0.0009  \ \ &  \ \ -0.2923 \ \ &  \ \ 4.0749 \ \ & \ \ -4.3085\ \ \ \\
\hline\hline
\end{tabular}\caption{The Wilson coefficients $C_{i}^{\mu}$ at the scale $\mu\sim m_{b}$ in the SM.}
\label{wc table}
\end{table*}

In the next subsection we give a brief review of different NP-scenarios\cite{DL,Alok:2017jgr,Capdevila:2017bsm} which will be used to analyze the physical observables of Rare $B\to K_{1}(1270,1400)\mu^{+}\mu^{-}$ decay.
\\
In SM and in NP, the effective Hamiltonian (\ref{H1})
gives the matrix element for $B\to K_{1}(1270,1400)\mu^{+}\mu^{-}$ can be written as
\begin{eqnarray}
\mathcal{M}(B\rightarrow K_{1} \mu^{+}\mu^{-})&=&\frac{G_F\alpha}{2\sqrt{2}\pi}V_{tb}V_{ts}^{\ast}
\bigg[\langle K_{1}(k,\varepsilon)|\overline{s}\gamma^{\mu}(1-\gamma^5)b|B(p)\rangle
\left\{C_9^{\text{tot}}(\overline{\mu}\gamma^{\mu}\mu)+C_{10}^{\text{tot}}(\overline{\mu}\gamma^{\mu}\gamma^5\mu)\right\}\notag \\
&&-2C_7^{eff} m_b\langle
K_{1}(k,\varepsilon)|\overline{s}i\sigma_{\mu\nu}\frac{q^{\nu}}{s}(1+\gamma^5)b|B(p)\rangle(\overline{\mu}\gamma^{\mu}\mu)\bigg],\label{Amplitude}
\end{eqnarray}

In the next subsection we give a brief review of different NP-scenarios\cite{DL,Alok:2017jgr,Capdevila:2017bsm} which will be used to analyze the physical observables of rare $B\to K_{1}(1270,1400)\mu^{+}\mu^{-}$ decay.
\subsection{New Physics Scenarios}
From the model-independent analysis performed in Ref.~\cite{Capdevila:2017bsm}, only the following three NP scenarios for $b\to s\mu^+\mu^-$ are allowed by the experimental data assuming real Wilson coefficients:
\begin{eqnarray}
  (I) &C^{\mu\mu}_{9}(\text{NP})<0\,,\notag\\
  (II)&C^{\mu\mu}_{9}(\text{NP})=-C^{\mu\mu}_{10}(\text{NP})<0\label{Cs}\,,\\
  (III)&C^{\mu\mu}_{9}(\text{NP})=-C^{\mu\mu\prime}_{9}(\text{NP})<0\,.\notag
\end{eqnarray}
Both scenarios (I) and (II) takes part to investigate the status of NP. However the scenario III is rejected because it predicts
$R_{K}=1$ and it disagrees with the experiment. The simplest possible NP models involve the tree-level exchange of leptoquark(LQ)
or $Z^{\prime}$ boson. It was shown in Ref.~\cite{DL} that scenario II can arise in LQ or $Z^{\prime}$-models, but scenario I is
only possible with a $Z^{\prime}$. The details containing LQ's and $Z^{\prime}$ are given in references\cite{CCO,BG,GH,SS,NF,IA}.
\subsection{Review of the fitting results for the NP Wilson coefficients}
\subsubsection{Model-independent scenarios}
In this work we use the Wilson coefficients fitted in \cite{DL} to make predictions for the unpolarized and polarized ratios $R_{K_1^{(L,T)}}$. Following the terms in \cite{DL},
fit-A was obtained using only $CP$-conserving $b\to s\mu^{+}\mu^{-}$ observables and fit-B using both the CP-conserving observables and $R_{K^{(\ast)}}$. The NP in both fit A and fit B can be
accommodated with the Wilson coefficients (WC's) $C^{\mu\mu}_{9}(\text{NP})$ and $C^{\mu\mu}_{10}(\text{NP})$ and the numerical values of these
WC's obtained in \cite{DL} are depicted in Table~\ref{fitMI}.
\begin{table}[tbh]
\centering \caption{Model-independent scenario: best-fit
values of the  WCs (taken to be real) as well as the pull values in fit-A and fit-B\cite{DL}}
\label{fitMI}%
\begin{tabular}{c|cc|cc}
\hline\hline
$\text{Scenario}$ & $\text{WC : fit-A}$ & $\text{pull}$ & $\text{WC : fit-B}$ & $\text{pull}$ \\ \hline
(I) $C^{\mu\mu}_{9}(\text{NP})$ & $ -1.20\pm 0.20$ & $ 5.0$ & $ -1.25\pm 0.19$ & $ 5.9$\\
(II) $C^{\mu\mu}_{9}(\text{NP})=-C^{\mu\mu}_{10}(\text{NP})$ & $ -0.62\pm 0.14$ & $ 4.6$ & $ -0.68\pm 0.12$ & $ 5.9$\\
(III)  $C^{\mu\mu}_{9}(\text{NP})=-C^{\mu\mu\prime}_{10}(\text{NP})$ & $ -1.10\pm 0.18$ & $ 5.2$ & $ -1.11\pm 0.17$ & $ 5.6$
 \\ \hline\hline
\end{tabular}
\end{table}

The simplest NP models that can explain the  $b\to s\mu^{+}\mu^{-}$ anomalies are the tree-level exchange of a new particle such as
a leptoquark (LQ) or a $Z^{\prime}$ boson. The details of the LQ and the $Z^{\prime}$ models were presented in \cite{DL,Alok:2017jgr} and
references therein. However in the next section we briefly discuss those leptoquark and $Z^{\prime}$ models that can explain the $b\to s\mu^{+}\mu^{-}$ data.
\subsubsection{Leptoquark models}
There are ten versions of leptoquarks that couple to the SM particles through dimension$\leq 4$ effective operators~\cite{AGC}. Among them, the scalar isotriplet $S_3$, the vector isosinglet $U_1$ and the vector isotriplet $U_3$ respectively with $Y=1/3, Y=-2/3$ and $Y=-2/3$ can explain the $b\to s\mu^{+}\mu^{-}$ data \cite{Alok:2017jgr,Sakaki:2013bfa} (and the $U_1$ can also simultaneously explain the $b\to c$ anomalies \cite{Bhattacharya:2016mcc,Bhattacharya:2014wla,Buttazzo:2017ixm,Angelescu:2018tyl}). The NP in LQ models can be accommodated via the Wilson coefficients,
$C_{9}^{\mu\mu}(NP)=-C_{10}^{\mu\mu}(NP)$. Such type of LQ models fall within the model-independent scenario II given in Eq.(\ref{Cs}), thus the
best-fit WC's for such LQ models remain the same as those for scenario II presented in Table~\ref{fitMI}. In these models LQ's are generated at the tree level and can be written as
\begin{eqnarray}
   C_{9}^{\mu\mu}(NP)\propto\frac{g_{L}^{b\mu}g_{L}^{s\mu}}{M^{2}_{LQ}}\,,\label{WC1}
\end{eqnarray}
where $g^{b\mu}_{L}$ and $g^{s\mu}_{L}$ are the couplings of the LQ and $M_{LQ}$ is the LQ mass, on which the constraint from the direct search is $M_{LQ}>640$ GeV~\cite{Aad:2015caa}.
\subsubsection{$Z^{\prime}$ models}
Like the variety of LQ models, there are also different versions of  $Z^{\prime}$ models \cite{Alok:2017jgr,CCO,Gauld:2013qba}. As discussed in previous sections, the $Z^{\prime}$ models that can explain the $b\to s\mu^+\mu^-$ anomalies should satisfy the model-independent scenarios I and II. Unlike the leptoquark models which can only be accommodated with scenario II, both I and II can be realized with a $Z^\prime$ exchange. To be general, in \cite{DL} both the heavy and light $Z^{\prime}$ models were considered. Next we briefly review their results.
 \\
 \\
 \textbf{(A). Heavy $Z^{\prime}$}\\
For $Z^{\prime}$ models of scenario I and II, by integrating out the heavy $Z^{\prime}$, the effective Lagrangian can be written as
\begin{eqnarray}
\mathcal{L}^{\textbf{eff}}_{Z^{\prime}}=-\frac{1}{2M^{2}_{Z^{\prime}}}J_{\mu}J^{\mu}\label{Lag1}
\end{eqnarray}
where
\begin{eqnarray*}
 J^{\mu}=-g^{\mu\mu}_{LL}\bar{L}\gamma^{\mu}P_{L}L+g^{\mu\mu}_{R}\bar{\mu}\gamma^{\mu}P_{R}\mu+g^{bs}_{L}\bar{\psi}_{q_{2}}\gamma^{\mu}P_{L}\psi_{q_{3}}+H.C.
\end{eqnarray*}
In terms of four fermion operators the effective Lagrangian(\ref{Lag1}) can be expressed as
\begin{eqnarray}
  \mathcal{L}^{\textbf{eff}}_{Z^{\prime}}=-\frac{g^{bs}_{L}}{M^{2}_{Z^{\prime}}}(\bar{s}\gamma^{\mu}b)(\bar{\mu}\gamma^{\mu}(g^{\mu\mu}_{L}P_{L}+g^{\mu\mu}_{R}P_{R})\mu)\,,\notag\\
 -\frac{(g^{bs}_{L})^{2}}{2M^{2}_{Z^{\prime}}}(\bar{s}\gamma^{\mu}P_{L}b)(\bar{s}\gamma^{\mu}P_{L}b)\,,\notag\\
 -\frac{g^{\mu\mu}_{L}}{M^{2}_{Z^{\prime}}}(\bar{\mu}\gamma^{\mu}(g^{\mu\mu}_{L}P_{L}+g^{\mu\mu}_{R}P_{R})\mu)
 (\bar{\nu}_{\mu}\gamma^{\mu}P_{L}\nu_{\mu})\,.\label{Leff}
\end{eqnarray}
In Eq.(\ref{Leff}) the first four-fermion operator is relevant to $b\to s\mu^{+}\mu^{-}$ transitions, the second operator makes contribution to the
$B^{0}_{s}-\bar{B}^{0}_{s}$ mixing and the third operator has effects on the neutrino trident production.

The NP effects in $Z^{\prime}$ models can modify the WC's
\begin{eqnarray}
 C_{9}^{\mu\mu}(\textbf{NP})=-\left[\frac{\pi}{\sqrt{2} G_{F}\alpha V_{tb}V^{\ast}_{ts}}\right]\frac{g^{bs}_{L}(g^{\mu\mu}_{L}
 +g^{\mu\mu}_{R})}{M^{2}_{Z^{\prime}}}\,,\notag\\
 C_{10}^{\mu\mu}(\textbf{NP})=\left[\frac{\pi}{\sqrt{2}G_{F}\alpha V_{tb}V^{\ast}_{ts}}\right]\frac{g^{bs}_{L}(g^{\mu\mu}_{L}
 -g^{\mu\mu}_{R})}{M^{2}_{Z^{\prime}}}\,.\label{C9ZP}
\end{eqnarray}
Considering the constraints from the $B^{0}_{s}-\bar{B}^{0}_{s}$ mixing and the neutrino trident production, the best-fit values of the couplings $g^{bs}_{L}$ and $g^{\mu\mu}_{L,R}$ were obtained in \cite{DL}, which are presented in Table~\ref{Tab3} and Table~\ref{Tab4}.
 \begin{table}[tbh]
\centering \caption{TeV $Z^{\prime}$ model (scenario I and II); best-fit values of $g^{bs}_{L}$ in fit-A \cite{DL}.}
\label{Tab3}%
\begin{tabular}{c|cc|cc}
\hline\hline
 \multicolumn{5}{c}{${M_{Z^{\prime}}}=1~\textbf{TeV}$} \\ \hline
 $g^{\mu\mu}_{L}$ & $Z^{\prime}(I):g^{bs}_{L}\times 10^{3}$ & $\text{pull}$ & $Z^{\prime}(II):g^{bs}_{L}\times 10^{3}$ & $\text{pull}$ \\ \hline
 $0.5$ & $-1.8\pm 0.3$ & $5.0$ & $-1.9\pm 0.4$ & $4.6$
 \\ \hline\hline
\end{tabular}
\end{table}
\begin{table}[tbh]
\centering \caption{TeV $Z^{\prime}$ model (scenario I and II); best-fit values of $g^{bs}_{L}$ in fit-B \cite{DL}.}
\label{Tab4}%
\begin{tabular}{c|cc|cc}
\hline\hline
 \multicolumn{5}{c}{${M_{Z^{\prime}}}=1~\textbf{TeV}$}\\ \hline
 $g^{\mu\mu}_{L}$ & $Z^{\prime}(I):g^{bs}_{L}\times 10^{3}$ & $\text{pull}$  & $Z^{\prime}(II):g^{bs}_{L}\times 10^{3}$ & $\text{pull}$\\ \hline
 $0.5$ & $-1.9\pm 0.3$ & $5.9$ & $-2.1\pm 0.4$ & $5.9$
 \\ \hline\hline
\end{tabular}
\end{table}
\\
\\
\textbf{(B). Light $Z^{\prime}$}

A light $Z^{\prime}$ is also possible to address the $b\to s\mu^{+}\mu^{-}$ data. Given the absence of any signature for such a state in the dimuon invariant mass, two typical $Z^\prime$ masses, $M_{Z^{\prime}}=10~\rm{GeV}>m_B$  and $M_{Z^{\prime}}=200~\rm{MeV}<2m_\mu$ can be considered. The corresponding $Z^{\prime}$ models are called the GeV $Z^{\prime}$ model and the MeV $Z^{\prime}$ model, which respectively has intimation for dark matter \cite{Cline:2017lvv} and nonstandard neutrino interactions \cite{Datta:2017pfz}. The MeV $Z^{\prime}$ model can also explain the muon $g-2$ \cite{Datta:2017pfz} \footnote{Recently in \cite{Raby:2017igl} it was shown that a heavier family
specific $Z^\prime$ may also resolve the muon g-2 anomaly if the chirality flipping effects are carefully considered.}.

For the light $Z^{\prime}$ models, the vertex $\bar{s}bZ^{\prime}$ takes the following form \cite{DL}
\begin{eqnarray}
 F(q^{2})\bar{s}\gamma^{\mu}P_{L}bZ^{\prime}_{\mu}\,,\label{ver}
\end{eqnarray}
where for $q^{2}<<m^{2}_{B}$, the form factor $F(q^{2})$ can be expanded as
\begin{eqnarray}
 F(q^{2})=a^{bs}_{L}+g^{bs}_{L}\frac{q^{2}}{m^{2}_{B}}+....\,.\label{Fq}
\end{eqnarray}
However the GeV $Z^{\prime}$ model is independent of the form factors, and the vertex factor
in this model is $a^{bs}_{L}$ for all $q^{2}$, while for the MeV $Z^{\prime}$ model, $a_L^{bs}$ is severely constrained by $B\to K\nu\bar\nu$ and thus can be neglected.
The modified WC's of the MeV and the GeV $Z^{\prime}$ models for $b\to s\mu^+\mu^-$ are
 \begin{eqnarray}
  C_{9}^{\mu\mu}(\textbf{NP})=\left[\frac{\pi}{\sqrt{2}G_{F}\alpha V_{tb}V_{ts}^{\ast}}\right]\times
  \frac{(a^{bs}_{L}+g^{bs}_{L}(q^{2}/m^{2}_{B}))(g^{\mu\mu}_{L}+g^{\mu\mu}_{R})}{q^{2}-M^{2}_{Z^{\prime}}}\,,\notag\\
  C_{10}^{\mu\mu}(\textbf{NP})=-\left[\frac{\pi}{\sqrt{2}G_{F}\alpha V_{tb}V_{ts}^{\ast}}\right]\times
  \frac{(a^{bs}_{L}+g^{bs}_{L}(q^{2}/m^{2}_{B}))(g^{\mu\mu}_{L}-g^{\mu\mu}_{R})}{q^{2}-M^{2}_{Z^{\prime}}}\,.\label{WC2}
 \end{eqnarray}
where for the GeV $Z^{\prime}$ model the numerical values of the coupling $a_L^{bs}$ are given in Table~\ref{Tab5}, which are obtained from fits to the $b\to s\mu^+\mu^-$ data with consideration of the constraint on $a_L^{bs}$ from $B^{0}_{s}-\bar{B}^{0}_{s}$ mixing. For the MeV $Z^\prime$ model, $g_L^{\mu\mu}=10^{-3}$ and $g_L^{bs}=2.1\times10^{-5}$ can be obtained for scenario I from the the neutrino trident production constraint plus a fit to the $b\to s\mu^+\mu^-$ data, with $\rm{pull}=4.4$.
 \begin{table}[tbh]
\centering \caption{ GeV $Z^{\prime}$ model (scenario I and II); best-fit values of $a^{bs}_{L}$ in fit-A \cite{DL}.}
\label{Tab5}%
\begin{tabular}{c|cc|cc}
\hline\hline
 \multicolumn{5}{c}{${M_{Z^{\prime}}}=10~\textbf{GeV}$} \\ \hline
 $g^{\mu\mu}_{L}\times 10^{2}$ & $Z^{\prime}(I):a^{bs}_{L}\times 10^{6}$  & $\text{pull}$ & $Z^{\prime}(II):a^{bs}_{L}\times 10^{6}$ & $\text{pull}$\\ \hline
 $1.2$ & $-5.2\pm 1.2$ & $4.2$ & $-7.2\pm 1.8$ & $4.5$
 \\ \hline\hline
\end{tabular}
\end{table}
\section{Form factors and mixing of $ K_{1}(1270)-K_{1}(1400)$}\label{sec3}
The exclusive $B\to K_{1}(1270,1400)\mu^{+}\mu^{-}$ decays involve the hadronic matrix elements of quark operators, which can be parameterized in terms of the form factors as
\begin{eqnarray}
 \langle K_{1}(k,\epsilon)|V_{\mu}|B(p)\rangle&=&\varepsilon^{\ast}_{\mu}(M_{B}+M_{K_{1}})V_{1}(q^{2})-
 (p+k)_{\mu}(\varepsilon^{\ast}.q)\frac{V_{2}(q^{2})}{M_{B}+M_{K_{1}}}\notag\\
 &-&q_{\mu}(\varepsilon^{\ast}.q)\frac{2M_{K_{1}}}{q^{2}}[V_{3}(q^{2})-V_{0}(q^{2})]\,,\label{F1}\\
 \langle K_{1}(k,\epsilon)|A_{\mu}|B(p)\rangle&=& -\frac{2i\epsilon_{\mu\nu\alpha\beta}}{M_{B}+M_{K_{1}}}
 \varepsilon^{\ast\nu}p^{\alpha}k^{\beta}A(q^{2})\,,\label{F2}
\end{eqnarray}
where $V^{\mu}=\bar s\gamma^{\mu}b$ and $A^{\mu}=\bar s\gamma^{\mu}\gamma^{5}b$ are vector and axial vector currents, $\varepsilon^{\ast\nu}$
are polarization vector of the axial vector meson. The relation for vector form factors in Eq.(\ref{F1}) are
\begin{eqnarray}
 V_{3}(q^{2})&=&\frac{M_{B}+M_{K_{1}}}{2M_{K_{1}}}V_{1}(q^{2})-\frac{M_{B}-M_{K_{1}}}{2M_{K_{1}}}V_{2}(q^{2})\,,\label{V3}\\
 V_{3}(0)&=&V_{0}(0)\,.\notag
\end{eqnarray}
The other contributions from the tensor form factors are
\begin{eqnarray}
 \langle K_{1}(k,\epsilon)|\bar si\sigma_{\mu\nu}q^{\nu}b|B(p)\rangle&=&[(M_{B}^{2}-M_{K_{1}}^{2})\varepsilon^{\ast}_{\mu}
 -(\varepsilon^{\ast}.q)(p+k)_{\mu}]T_{2}(q^{2})\notag\\
 &+&(\varepsilon^{\ast}.q)\left[q_{\mu}-\frac{q^{2}}{M^2_{B}-M^2_{K_{1}}}(p+k)_{\mu}\right]T_{3}(q^{2})\,,\label{T1}\\
 \langle K_{1}(k,\epsilon)|\bar si\sigma_{\mu\nu}q^{\nu}\gamma^{5}b|B(p)\rangle&=&2i\epsilon_{\mu\nu\alpha\beta}\varepsilon^{\ast\nu}p^{\alpha}k^{\beta}T_{1}(q^{2})\,.\label{T2}
\end{eqnarray}
The physical states $K_{1}(1270)$ and $K_{1}(1400)$ are mixed states of $K_{1A}$ and $K_{1B}$ with mixing angle $\theta_{K}$ defined as
\begin{eqnarray}
 |K_{1}(1270)\rangle&=&|K_{1A}\rangle\sin\theta_{K_{1}}+|K_{1B}\rangle\cos\theta_{K_{1}}\,,\label{theta}\\
 |K_{1}(1400)\rangle&=&|K_{1A}\rangle\cos\theta_{K_{1}}-|K_{1B}\rangle\sin\theta_{K_{1}}\,.\label{theta1}
\end{eqnarray}
In terms of $K_{1A}$ and $K_{1B}$, the matrix element $B\to K_{1}(1270,1400)$ can be parameterized in terms of the form factors as
\begin{eqnarray}
 \begin{pmatrix}
  \langle K_{1}(1270)|\bar s\gamma_{\mu}(1-\gamma_{5})b|B\rangle \\
  \langle K_{1}(1400)|\bar s\gamma_{\mu}(1-\gamma_{5})b|B\rangle
 \end{pmatrix}
 =M\begin{pmatrix}
  \langle K_{1A}|\bar s\gamma_{\mu}(1-\gamma_{5})b|B\rangle \\
  \langle K_{1B}|\bar s\gamma_{\mu}(1-\gamma_{5})b|B\rangle\label{M1}
 \end{pmatrix}
 \,,
\end{eqnarray}
\begin{eqnarray}
 \begin{pmatrix}
  \langle K_{1}(1270)|\bar s\sigma_{\mu\nu}q^{\nu}(1+\gamma_{5})b|B\rangle \\
  \langle K_{1}(1400)|\bar s\sigma_{\mu\nu}q^{\nu}(1+\gamma_{5})b|B\rangle
 \end{pmatrix}
 =M\begin{pmatrix}
  \langle K_{1A}|\bar s\sigma_{\mu\nu}q^{\nu}(1+\gamma_{5})b|B\rangle \\
  \langle K_{1B}|\bar s\sigma_{\mu\nu}q^{\nu}(1+\gamma_{5})b|B\rangle\label{M2}
 \end{pmatrix}
 \,,
\end{eqnarray}
where the mixing matrix $M$ can be written as
\begin{eqnarray}
M=
 \begin{pmatrix}
  \sin\theta_{K}&\cos\theta_{K}\\
  \cos\theta_{K}&-\sin\theta_{K}\label{M3}
 \end{pmatrix}
 \,.
\end{eqnarray}
 The form factors used in the analysis of physical observables were calcualted in the framework of QCD light cone sum rules. These results are applicable only at low $q^{2}$
 region. However to investigate the effects of observables on the whole kinematical region, the form factors can be parameterized
 in the three-parameter form as \cite{Hatanaka:2008gu}\footnote{The choice of the hadronic form factors makes very tiny difference in the analysis due to the fact that the form factors in the LFU ratios essentially cancel.}
\begin{equation}
\mathcal{T}^{X}_{i}(q^{2})=\frac{\mathcal{T}^{X}_{i}(0)}{1-a_{i}^{X}\left(q^{2}/m^{2}_{B}\right)+b_{i}^{X}\left(q^{2}/m^{2}_{B}\right)^{2}}\,,\label{m11}
\end{equation}
where $\mathcal{T}$ is $A$, $V$ or $F$ form factors and the subscript
$i$ can take a value 0, 1, 2 or 3, and the superscript $X$ denotes the
$K_{1A}$ or $K_{1B}$ states. The numerical results of form factors at $q^{2}=0$ are presented in Table-\ref{tabel1}.
\begin{table*}[tbp]
\begin{tabular}{p{.7in}p{.7in}p{.7in}p{.4in}|p{.7in}p{.7in}p{.7in}p{.4in}}
\hline \hline
$\mathcal{T}^{X}_{i}(q^{2})$&$\mathcal{T}(0)$&$a$&$b$&$\mathcal{T}^{X}_{i}(q^{2})$&$\mathcal{T}(0)$&$a$&$b$\\
\hline
$V_{1}^{K_{1A}}$&$0.34$&$0.635$&$0.211$&$V_{1}^{K_{1B}}$&$-0.29$&$0.729$&$0.074$\\
$V_{2}^{K_{1A}}$&$0.41$&$1.51$&$1.18$&$V_{1}^{K_{1B}}$&$-0.17$&$0.919$&$0.855$\\
$V_{0}^{K_{1A}}$&$0.22$&$2.40$&$1.78$&$V_{0}^{K_{1B}}$&$-0.45$&$1.34$&$0.690$\\
$A^{K_{1A}}$&$0.45$&$1.60$&$0.974$&$A^{K_{1B}}$&$-0.37$&$1.72$&$0.912$\\
$F_{1}^{K_{1A}}$&$0.31$&$2.01$&$1.50$&$F_{1}^{K_{1B}}$&$-0.25$&$1.59$&$0.790$\\
$F_{2}^{K_{1A}}$&$0.31$&$0.629$&$0.387$&$F_{2}^{K_{1B}}$&$-0.25$&$0.378$&$-0.755$\\
$F_{3}^{K_{1A}}$&$0.28$&$1.36$&$0.720$&$F_{3}^{K_{1B}}$&$-0.11$&$1.61$&$10.2$\\
\hline\hline
\end{tabular}
\caption{$B\to K_{1A,1B}$ form factors \cite{Hatanaka:2008gu}, where $a$ and $b$
are the parameters of the form factors in dipole parametrization.}
\label{tabel1}
\end{table*}
\section{LFU Ratios $R_{K^{(L,T)}_1}$ and Ratio $R_\mu(K_1)$}\label{sec4}
In this section we present the formalism for the LFU ratios $R_{K_{1}}$  and the ratio $R_\mu(K_1)$,
considering unpolarized and polarized (longitudinal and transverse) final state axial-vector mesons $K_{1}(1270,1400)$,
and their sensitivity for different NP scenarios (scenario I and scenario II) or NP models (leptoquark models and heavy and light $Z^\prime$ models).
The $R_{K_{1}}$ parameter is a good tool to investigate NP, as the form factors in this observable is almost cancels out.
We now define unpolarized and polarized LFU ratios as:
\begin{eqnarray}
 R_{K^{(L,T)}_{1}(1270,1400)}(q^2)=\frac{d\mathcal B(B\to K^{(L,T)}_{1}(1270,1400)\mu^{+}\mu^{-})/dq^{2}}{d\mathcal B(B\to K^{(L,T)}_{1}(1270,1400)e^{+}e^{-})/dq^{2}}.\label{RK1}
\end{eqnarray}
Since $K_{1}$ meson involves the mixing angle $\theta_{K_{1}}$, therefore to determine the mixing angle $\theta_{K_{1}}$, we define
another ratio $R_\mu$ for $K_1$ mesons as
\begin{eqnarray}
 R_\mu(K^{(L,T)}_{1})(q^2)=\frac{d\mathcal B(B\to K^{(L,T)}_{1}(1400)\mu^{+}\mu^{-})/dq^{2}}{d\mathcal B(B\to K^{(L,T)}_{1}(1270)\mu^{+}\mu^{-})/dq^{2}}.\label{RK1}
\end{eqnarray}
To compute the above ratios we use the amplitude for the $B\to K_{1}\mu^{+}\mu^{-}$ decays given in Eq.(\ref{Amplitude}).

The matrix element given in Eq.(\ref{Amplitude}) can also be written as
\begin{eqnarray}
 \mathcal{M}&=&\frac{G_{F}\alpha}{2\sqrt{2}\pi}\lambda_{t}\{T_{1}^{\mu}\bar{\mu}\gamma_{\mu}\mu+T_{2}^{\mu}\bar{\mu}\gamma_{\mu}\gamma_{5}\mu\}\label{AMP1}
\end{eqnarray}
where the form factors and Wilson coefficients are hidden in $T^{\mu}_{i}$ which can be expressed as follows:
\begin{eqnarray}
 T^{\mu}_{i}=T^{\mu\nu}_{i}\varepsilon^{\ast}_{\nu} (i=1,2)
\end{eqnarray}

and
\begin{eqnarray}
 {T_1}_{\mu\nu}&=&\{-i\epsilon_{\mu\nu\alpha\beta}
 p^{\alpha}k^{\beta}\mathcal{F}_{1}(q^{2})-g_{\mu\nu}\mathcal{F}_{2}(q^{2})+P_{\mu}q_{\nu}\mathcal{F}_{3}(q^{2})\notag\\
 &+&q_{\mu}q_{\nu}\mathcal{F}_{4}(q^{2})\},\label{M3}\\
 {T_2}_{\mu\nu} &=&\{-i\epsilon_{\mu\nu\alpha\beta}
 p^{\alpha}k^{\beta}\mathcal{F}_{5}(q^{2})-g_{\mu\nu}\mathcal{F}_{6}(q^{2})+P_{\mu}q_{\nu}\mathcal{F}_{7}(q^{2})\notag\\
 &+&q_{\mu}q_{\nu}\mathcal{F}_{8}(q^{2})\},,\label{M4}
\end{eqnarray}
where the auxiliary functions $\mathcal{F}_{1},........,\mathcal{F}_{8}$ accommodate both the form factors and Wilson coefficients. The explicit expressions for them can be written as
follows
\begin{eqnarray}
 \mathcal{F}_{1}(q^{2})&=&C_{9}^{\text{tot}}\frac{A(q^{2})}{M_{B}+M_{K_{1}}}+\frac{4m_{b}}{q^{2}}C_{7}^{\text{eff}}T_{1}(q^{2})\,,\notag\\
 \mathcal{F}_{2}(q^{2})&=& C_{9}^{\text{tot}}(M_{B}+M_{K_{1}})V_{1}(q^{2})+\frac{2m_{b}}{q^{2}}C_{7}^{\text{eff}}(M^{2}_{B}-M^{2}_{K_{1}})T_{2}(q^{2})\,,\notag\\
 \mathcal{F}_{3}(q^{2})&=&C_{9}^{\text{tot}}\frac{V_{2}(q^{2})}{M_{B}+M_{K_{1}}}-\frac{2m_{b}}{q^{2}}C_{7}^{\text{eff}}(T_{2}(q^{2})-\frac{q^{2}}{M^{2}_{B}-M^{2}_{K_{1}}}T_{3}(q^{2}))\,,\label{A1}\\
 \mathcal{F}_{4}(q^{2})&=&C_{9}^{\text{tot}}\frac{2M_{K_{1}}}{q^{2}}[V_{3}(q^{2})-V_{0}(q^{2})]-\frac{2m_{b}}{q^{2}}C_{7}^{\text{eff}}T_{3}(q^{2})\,,\notag
\end{eqnarray}
and also
\begin{eqnarray}
 \mathcal{F}_{5}(q^{2})&=&2C_{10}^{\text{tot}}\frac{A(q^{2})}{M_{B}+M_{K_{1}}}\,,\notag\\
 \mathcal{F}_{6}(q^{2})&=& C_{10}^{\text{tot}}(M_{B}+M_{K_{1}})V_{1}(q^{2})\,,\notag\\
 \mathcal{F}_{7}(q^{2})&=&C_{10}^{\text{tot}}\frac{V_{2}(q^{2})}{M_{B}+M_{K_{1}}}\,,\label{A2}\\
 \mathcal{F}_{8}(q^{2})&=&C_{10}^{\text{tot}}\frac{2M_{K_{1}}}{q^{2}}[V_{3}(q^{2})-V_{0}(q^{2})]\,.\notag
\end{eqnarray}
Since the final state mesons $K_{1}(1270)$ and $K_{1}(1400)$ involve the mixing angle $\theta_{K}$, the form factors of $K_1(1270,1400)$ in Eqs.(\ref{A1}) and (\ref{A2})
can be written in terms of the form factors of $K_{1A}$ and $K_{1B}$:\\
 \textbf{$B\to K_{1}(1270)$ form factors in terms of mixing angle $\theta_{K}$}
 \begin{eqnarray}
  A(q^{2})&=&-A^{K_{1A}}\sin\theta_{K}+A^{K_{1B}}\cos\theta_{K}\,,\notag\\
  V_{i}(q^{2})&=&-V_{i}^{K_{1A}}\sin\theta_{K}+V_{i}^{K_{1B}}\cos\theta_{K}\,,\label{K1(1270)}\\
  F_{i}(q^{2})&=&-F_{i}^{K_{1A}}\sin\theta_{K}+F_{i}^{K_{1B}}\cos\theta_{K}\,.\notag
 \end{eqnarray}
 \textbf{$B\to K_{1}(1400)$ form factors in terms of mixing angle $\theta_{K}$}
 \begin{eqnarray}
  A(q^{2})&=&A^{K_{1A}}\cos\theta_{K}-A^{K_{1B}}\cos\theta_{K}\,,\notag\\
  V_{i}(q^{2})&=&V_{i}^{K_{1A}}\cos\theta_{K}-V_{i}^{K_{1B}}\sin\theta_{K}\,,\label{K1(1400)}\\
  F_{i}(q^{2})&=&F_{i}^{K_{1A}}\cos\theta_{K}-F_{i}^{K_{1B}}\sin\theta_{K}\,.\notag
 \end{eqnarray}
\section{Phenomenological Analysis of $R_{K_1^{(L,T)}}$ and $R_\mu(K_1)$}\label{sec5}
In this section, we give predictions for the unpolarized and polarized LFU ratios $R_{K_1^{(L,T)}}$ \footnote{Comparison of $R_{K^{(*)}}$, $R_{K_0}$ and $R_{K_1}$ and detailed discussions based on symmetries can be found in \cite{Hiller:2014ula}.} in the SM and in the NP models under consideration. In the numerical calculation, we adopt the following input parameters\cite{pdg}:
\begin{gather*}
m_{B}=5.28~{\rm GeV},~~ m_{b}=4.18~{\rm GeV}, ~~m_{\mu}=0.105~{\rm GeV},\\
m_{\tau}=1.77~{\rm GeV},~~ f_{B}=0.25~{\rm GeV}, ~~|V_{tb}V_{ts}^{\ast}|=41\times10^{-3},\\
\alpha^{-1}=137,~~ G_{F}=1.17\times 10^{-5} {\rm GeV^{-2}},~~ \tau_{B}=1.54\times 10^{-12} s,\\
m_{K_{1}(1270)}=1.270~{\rm GeV},~~ m_{K_{1}(1400)}=1.403~{\rm GeV}.
\end{gather*}
To obtain the results in the model-independent scenarios, the Wilson coefficients given in Table~\ref{fitMI} are used. As mentioned previously, the model-independent Wilson coefficients in Scenario II can also be achieved in the leptoquark models, therefore the corresponding predictions also represent the results in the leptoquark models. For $Z^{\prime}$ models, we obtain our predictions using the couplings listed in Table~\ref{Tab3}-\ref{Tab5}. We present the results obtained in different NP models and NP scenarios in separate plots, but one will see that the major factor that affects the predictions are the NP scenario I and II rather than the specific NP models which means the plots corresponding to the leptoquark, heavy and light $Z^\prime$ models are similar once they fulfill the same NP scenario. This is not surprising because from Table~\ref{fitMI}-\ref{Tab5} one sees the pull values for the same NP scenario (but in different NP models) are close, which means the leptoquark models and $Z^{\prime}$ models can reproduce or nearly reproduce the model-independent results under current experimental constraints. Furthermore, since the $K_1$ mixing angle has not been precisely determined, to be more general, we also consider different possibilities of the $K_1$ mixing angle \cite{Hatanaka:2008gu,Hatanaka:2008xj} in our analysis.

In the figures of this section we plot the physical observables in low and high $q^{2}$ regions, as we already discussed in Section \ref{sec2} the
functions $F^{7,9}_{8}(q^{2})$ and $F^{7,9}_{1c}(q^{2})$ involved in the definition of Wilson coefficents $C^{\text{eff}}_{7}(q^{2})$
and $C^{\text{eff}}_{7}(q^{2})$ given in Eq.(\ref{WC3}) defined for low and high $q^{2}$ separately.
To provide a comparison with future experimental results,  the LFU ratios in
low and high $q^2$ bins are presented in Appendix~\ref{appA}.
\begin{figure}[!htbp]
\begin{center}
\includegraphics[scale=0.6]{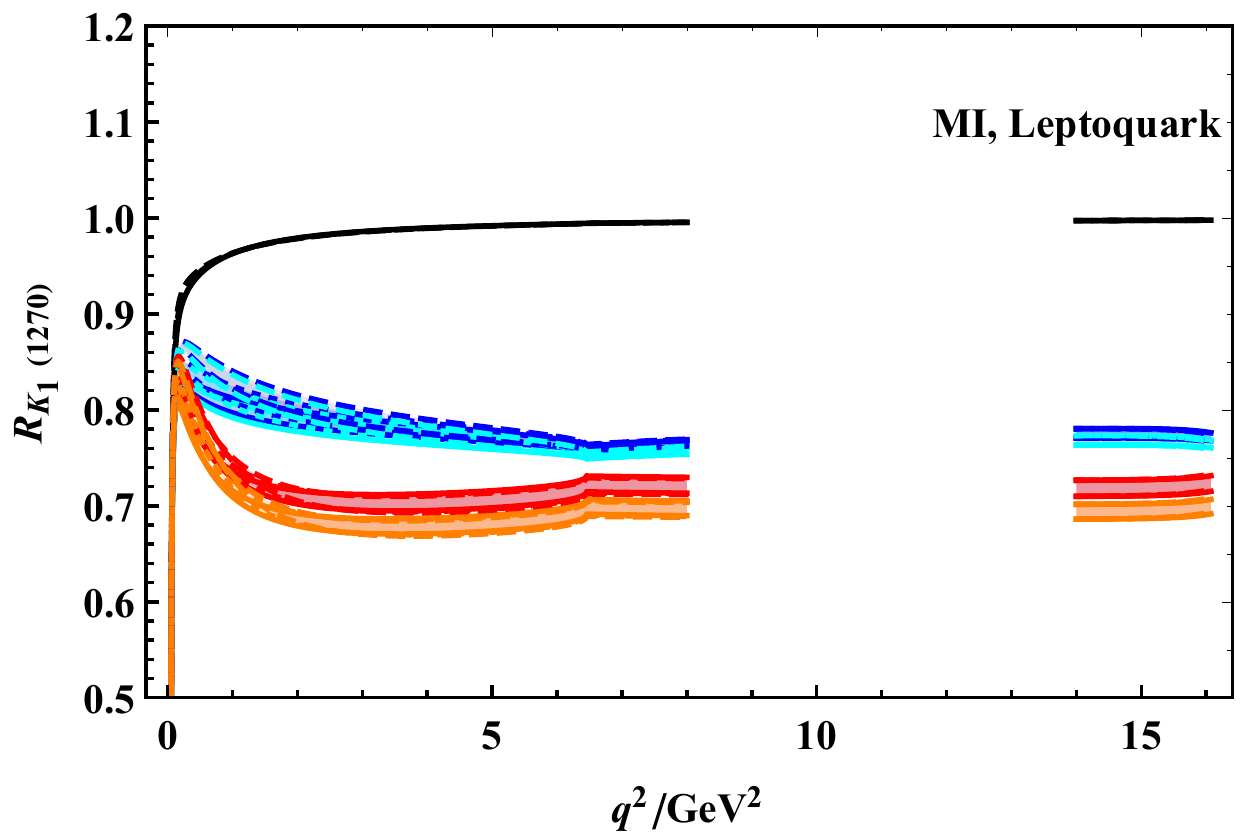}
\includegraphics[scale=0.6]{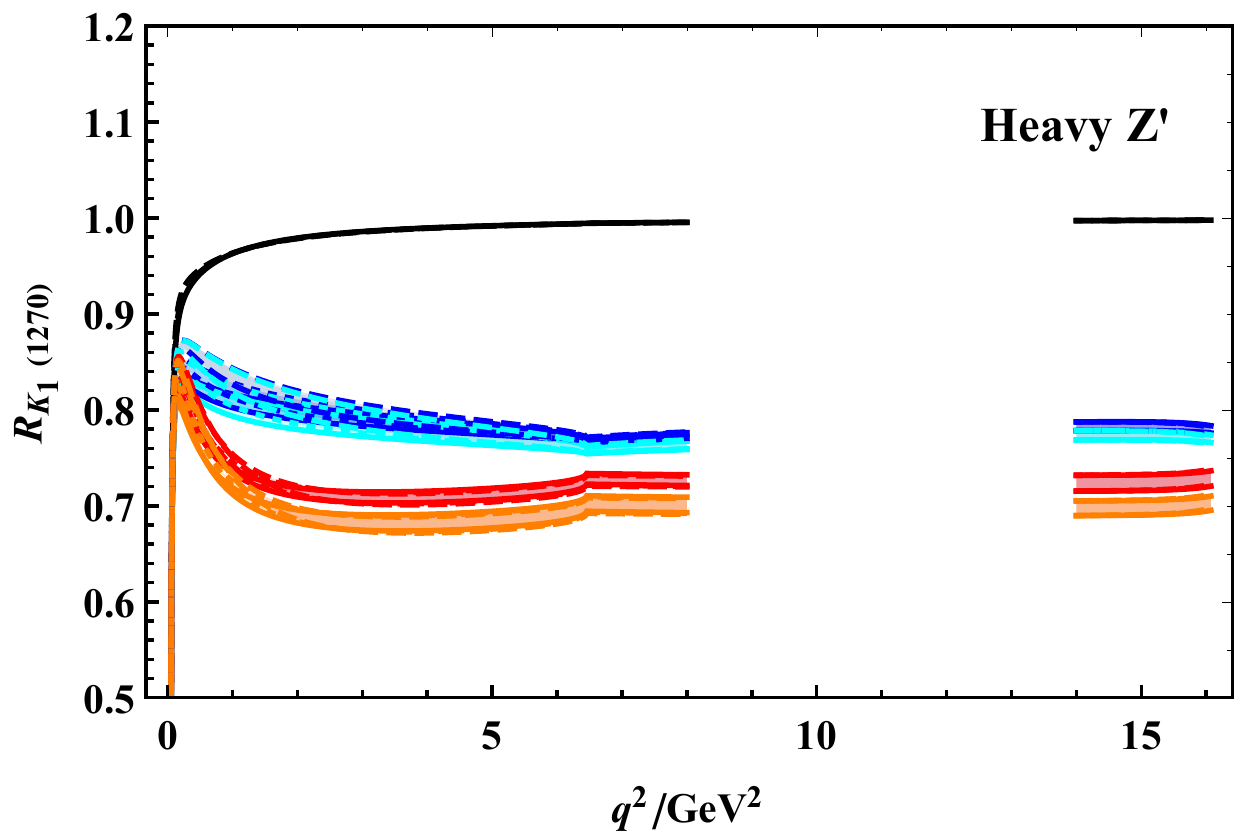}
\includegraphics[scale=0.6]{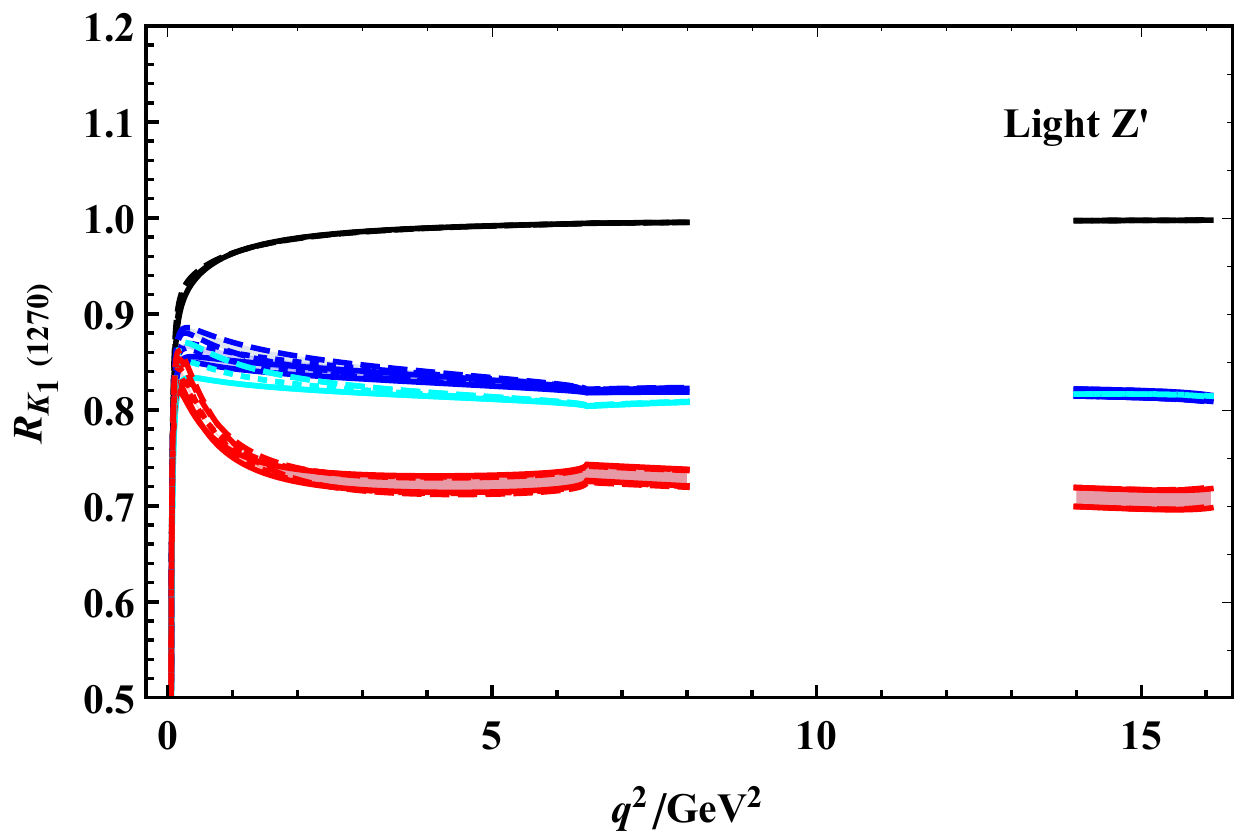}
\caption{The Standard Model, model-independent scenarios, leptoquark models, and heavy and light $Z^\prime$ models predictions for the LFU ratio $R_{K_1(1270)}$. The black curves denote the predictions in the SM, the blue, cyan, red and orange bands show the predictions obtained in scenario I(A), I(B), II(A) and II(B), respectively, including the errors due to errors of the modified Wilson coefficients. For each scenario, the bands with solid, dotted and dashed boundary curves correspond to $\theta_{K_1}=-\ang{34}, -\ang{45}, -\ang{57}$, respectively. The results in scenario II(A) and II(B) also represent the predictions from the leptoquark models.}
\label{fig:RK11270}
\end{center}
\end{figure}

In Fig.~\ref{fig:RK11270}, we have plotted the LFU parameter $R_{K_1(1270)}$ against the square of the momentum transfer $q^2$ in the SM and in different NP models under consideration. One can see that for a given $q^2$ region, the impact of the NP on this observable is distinct from the SM value which is $\approx1$. It can also be noticed that the value of $R_{K_1(1270)}$ in low $q^2$ region decreases when the value of $q^2$ increases.
However, in the region above $q^2=4$ GeV$^2$ the observable $R_{K_1(1270)}$ does not vary with the value of $q^2$.
This figure also shows that the variation in the values of $R_{K_1(1270)}$ due to the different NP models are almost the same.
However, as compared to the SM, the behavior of $R_{K_1(1270)}$ due to scenario I and scenario II are clearly distinguishable. This suggests that the precise measurement of $R_{K_1(1270)}$ in current and future colliders will segregate the SM from the leptoquark and the $Z^{\prime}$ models. Moreover if scenario I is observed,  it can only be realized in $Z^{\prime}$ models.
\begin{figure}[!htbp]
\begin{center}
\includegraphics[scale=0.6]{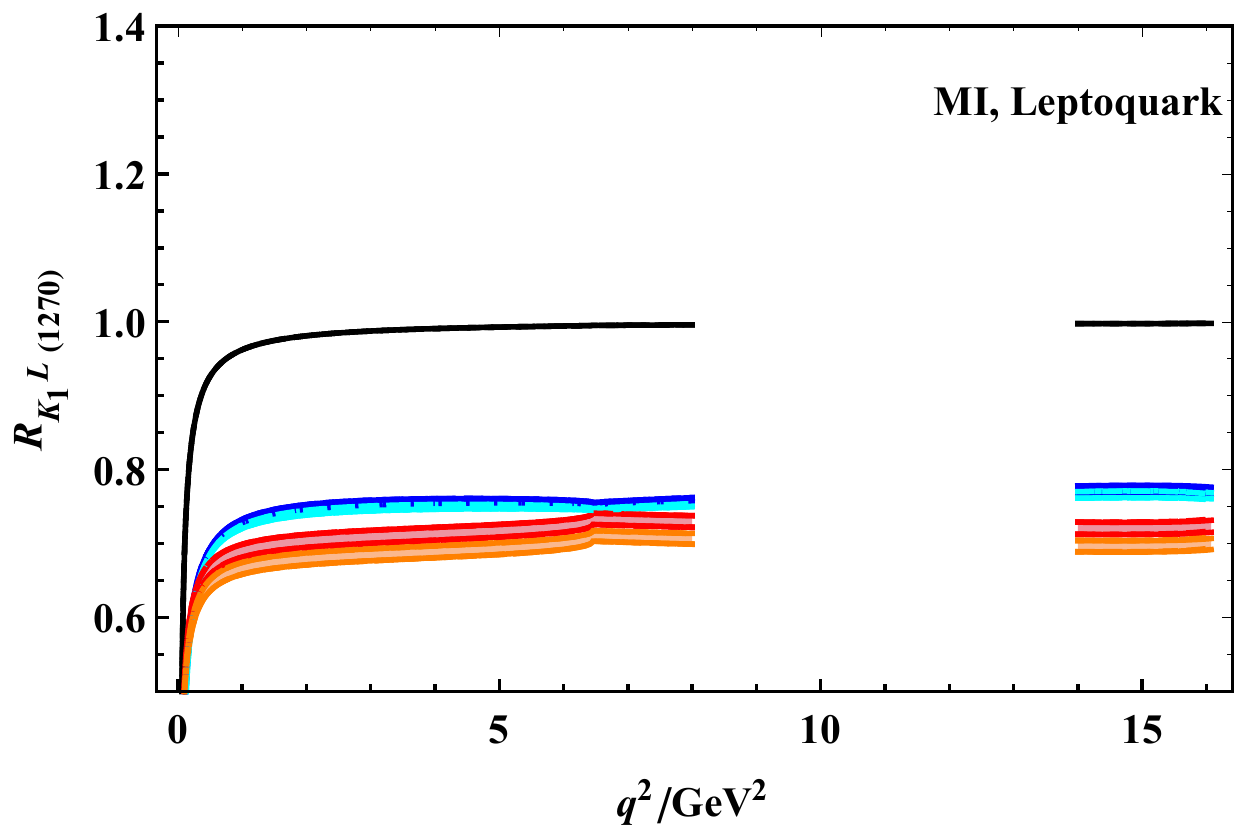}
\includegraphics[scale=0.6]{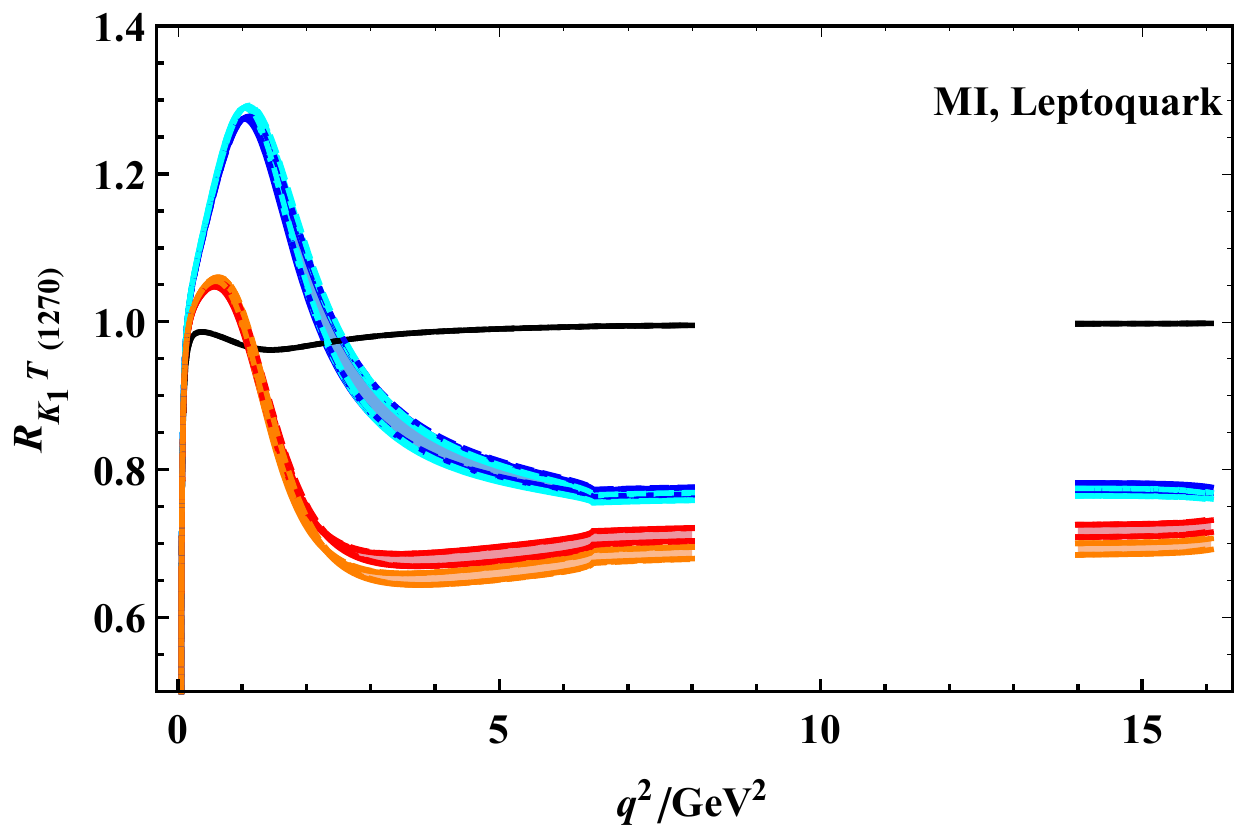}
\includegraphics[scale=0.6]{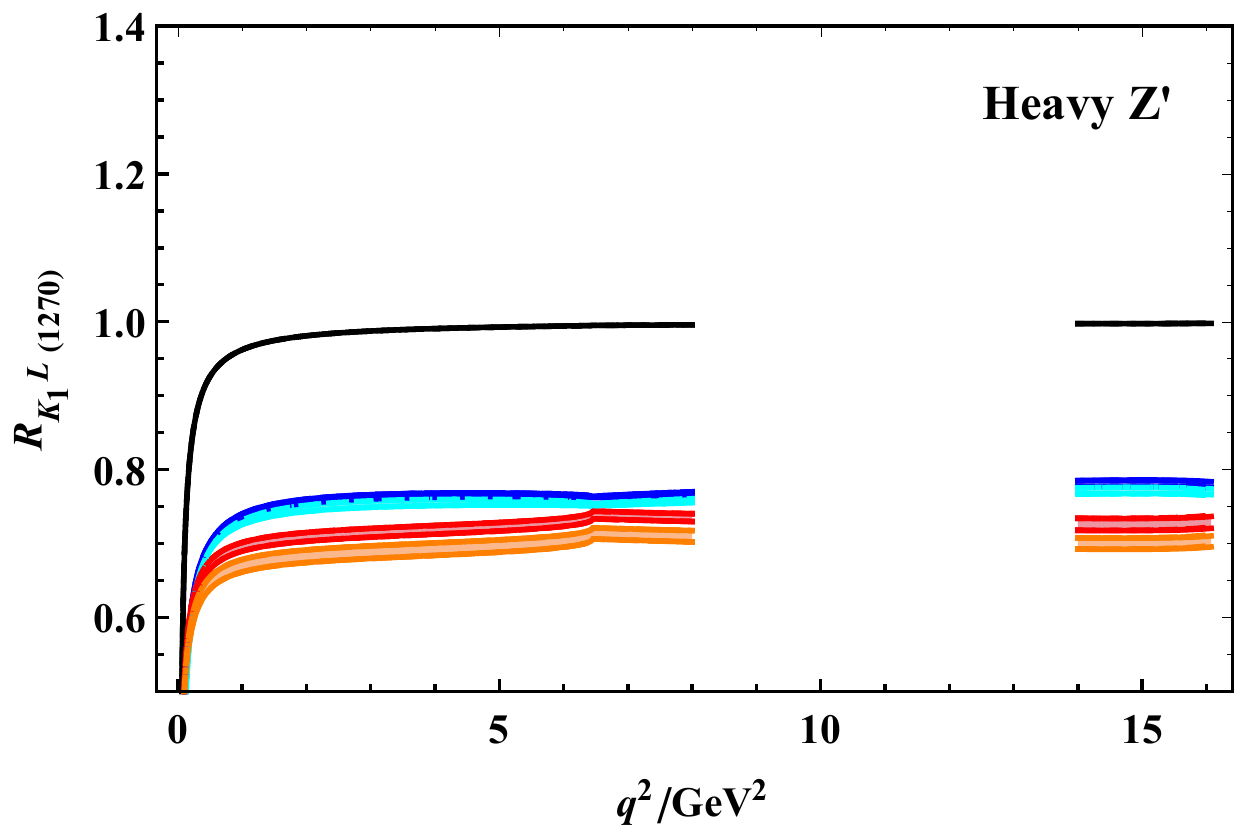}
\includegraphics[scale=0.6]{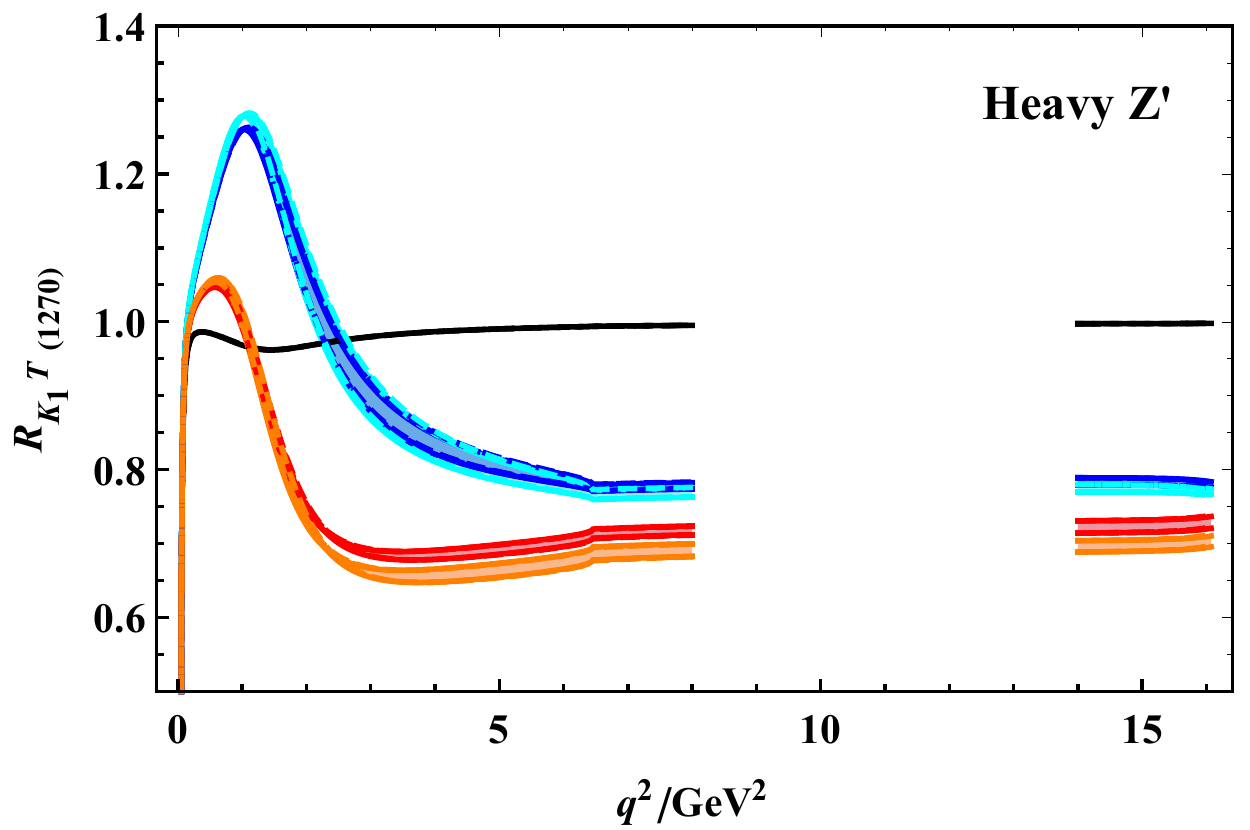}
\includegraphics[scale=0.6]{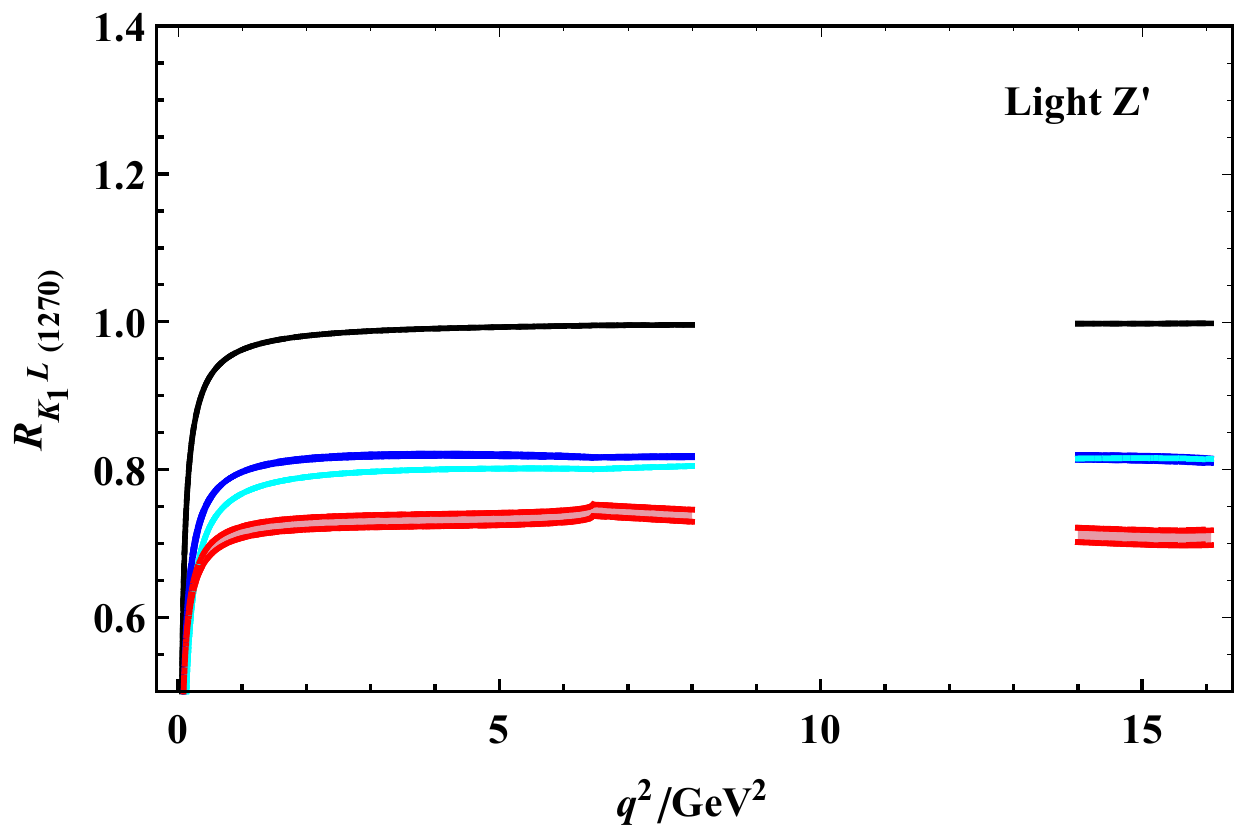}
\includegraphics[scale=0.6]{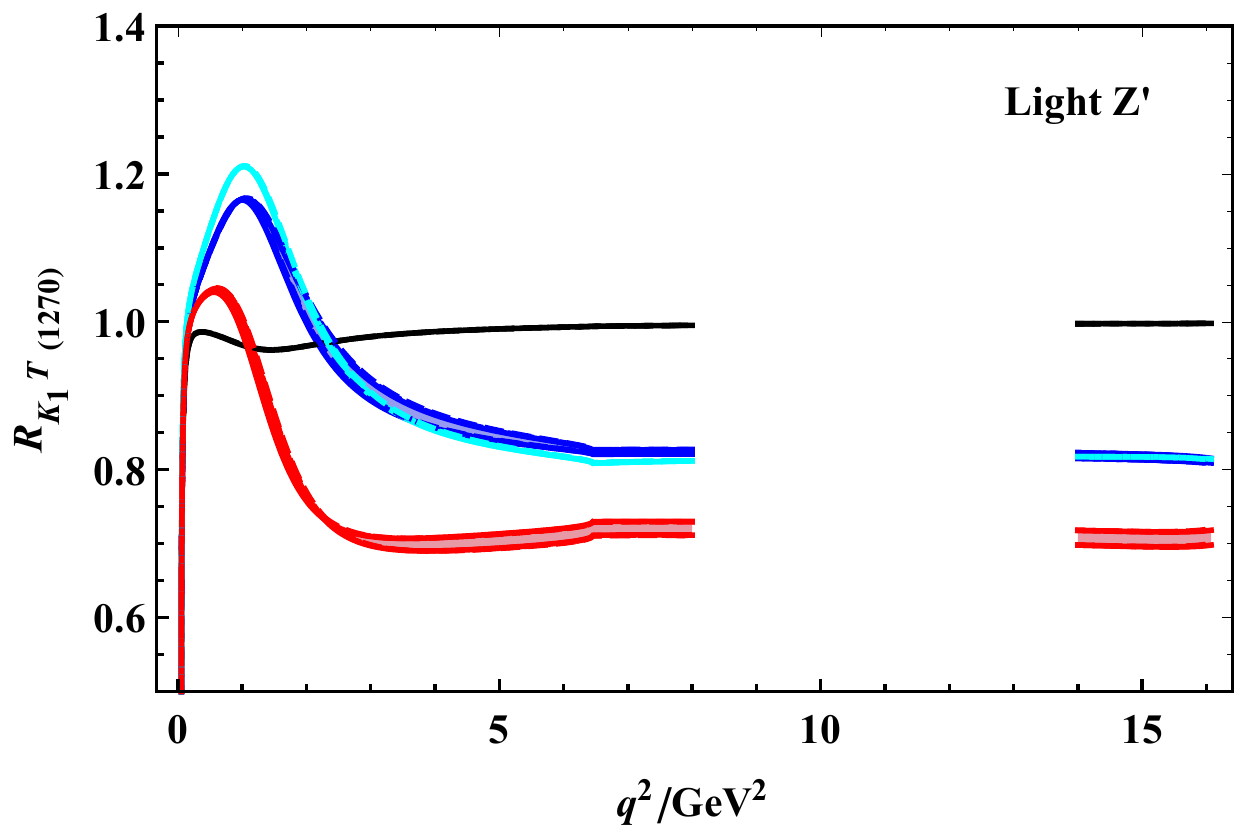}
\caption{Predictions for the polarized LFU ratios $R_{K_1^{(L,T)}(1270)}$ in the SM, model-independent scenarios, leptoquark, heavy and light $Z^\prime$ models. The legends are the same as in Fig.~\ref{fig:RK11270}.}
\label{fig:RK1p1270}
\end{center}
\end{figure}

Similarly, in Fig.~\ref{fig:RK1p1270}, we have plotted the polarized LFU parameters $R_{K^{L,T}_1(1270)}$
(i.e. the ratio when $K_1$ meson is longitudinally or transversely polarized) against the square of the momentum transfer $q^2$
in the SM and in the different NP models. This figure also represent that the NP effects are quite distinguishable. For the case of the longitudinal LFU parameter $R_{K^{L}_{1}(1270)}$ it is shown that by increasing the value of $q^{2}$, the behavior of
the observable $R_{K^{L}_{1}(1270)}$ remains stable in both scenario-I and scenario-II for all NP models under consideration.
However the values of $R_{K^{L}_{1}(1270)}$ in scenario-I and scenario-II are distinguishable and are approximately 0.75 to 0.80 and 0.65 to 0.70 respectively.
Furthermore on the right panel of Fig.~\ref{fig:RK1p1270} one can see the value of the transverse LFU parameter $R_{K^{T}_{1}(1270)}$
does vary at low $q^{2}$ region, i.e. around $q^{2}=1-2$ $\text{GeV}^{2}$, the value of $R_{K^{T}_{1}(1270)}$ exceeds 1 for all NP models under discussion.
Therefore, these polarized observables, particularly the $R_{K^{T}_1(1270)}$ in low $q^2$ region are useful to probe the effects of a leptoquark or a $Z^\prime$ Model.
\begin{figure}[!htbp]
\begin{center}
\includegraphics[scale=0.6]{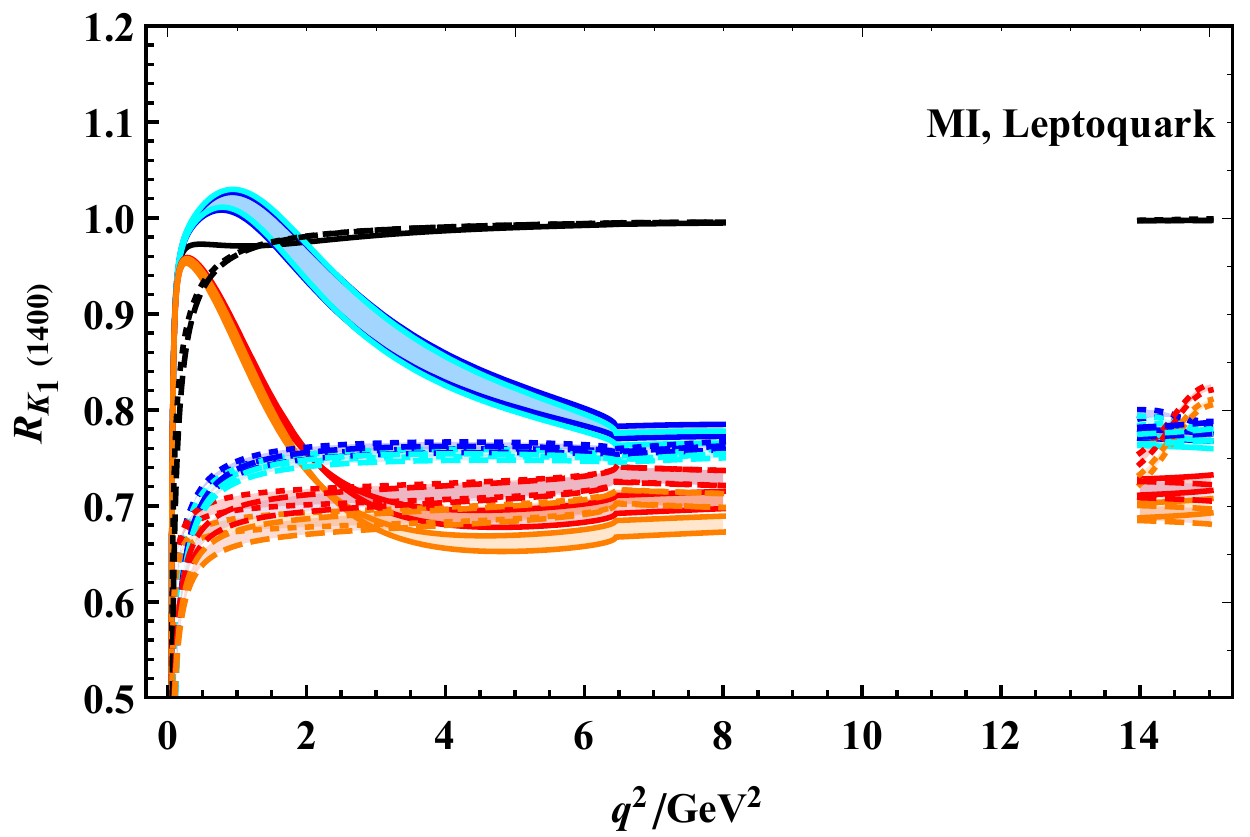}
\includegraphics[scale=0.6]{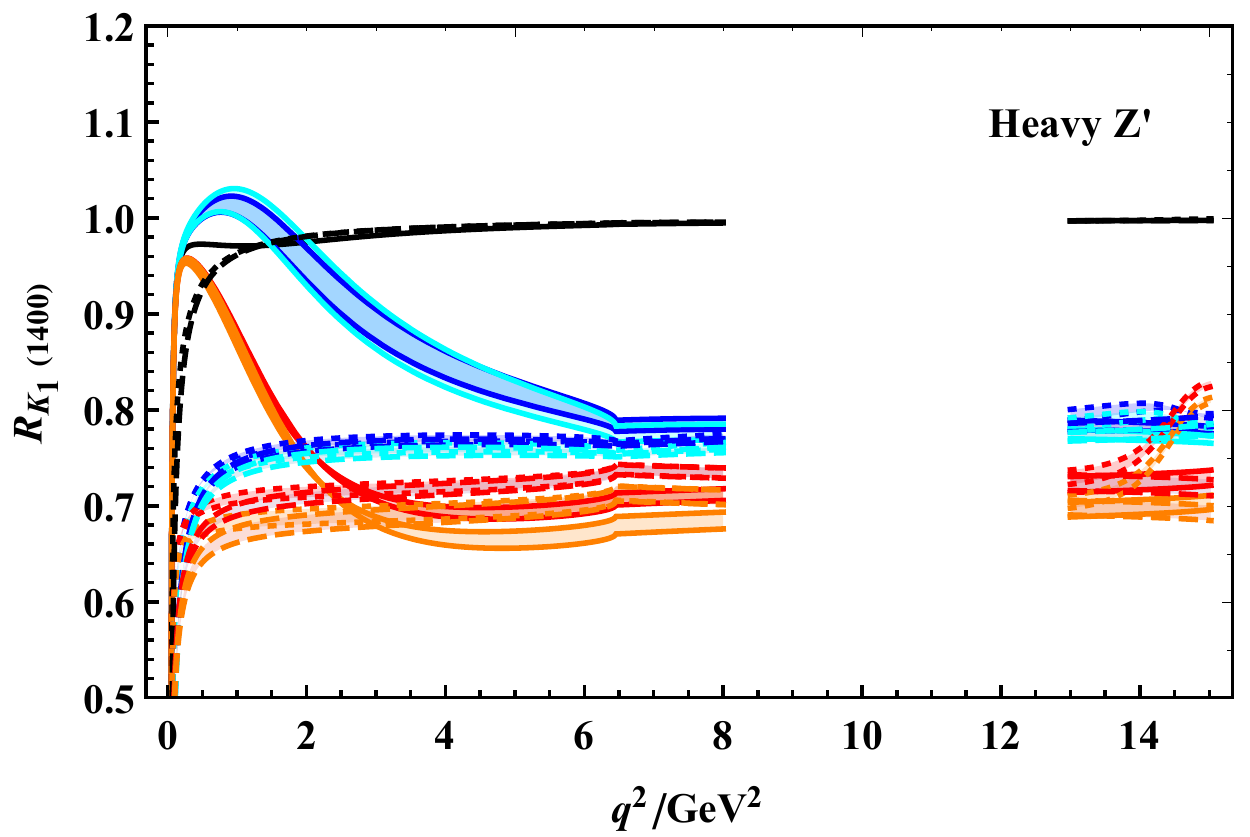}
\includegraphics[scale=0.6]{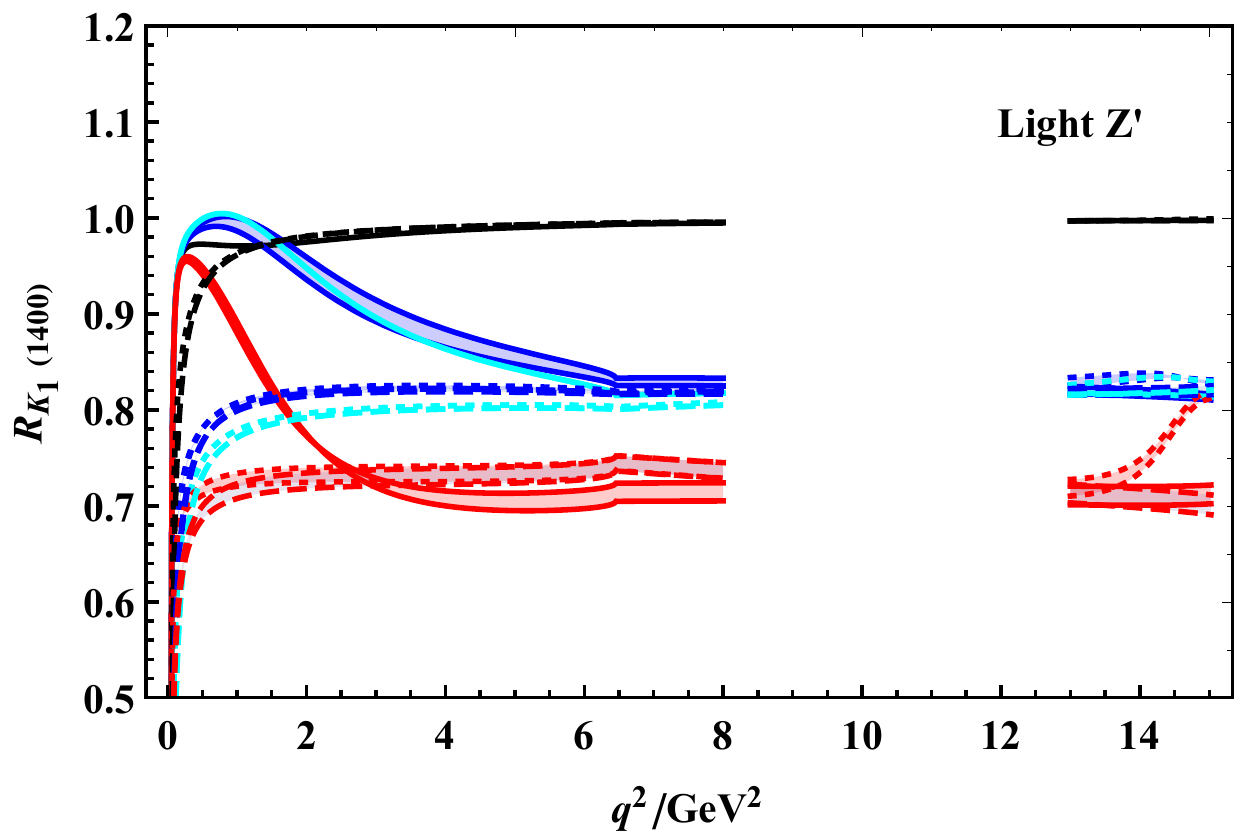}
\caption{Predictions for the polarized LFU ratios $R_{K_1(1400)}$ in the Standard Model, model-indepenent scenarios, leptoquark models, heavy and light $Z^\prime$ models. The legends are the same as in Fig.~\ref{fig:RK11270}.}
\label{fig:RK11400}
\end{center}
\end{figure}

\begin{figure}[!htbp]
\begin{center}
\includegraphics[scale=0.6]{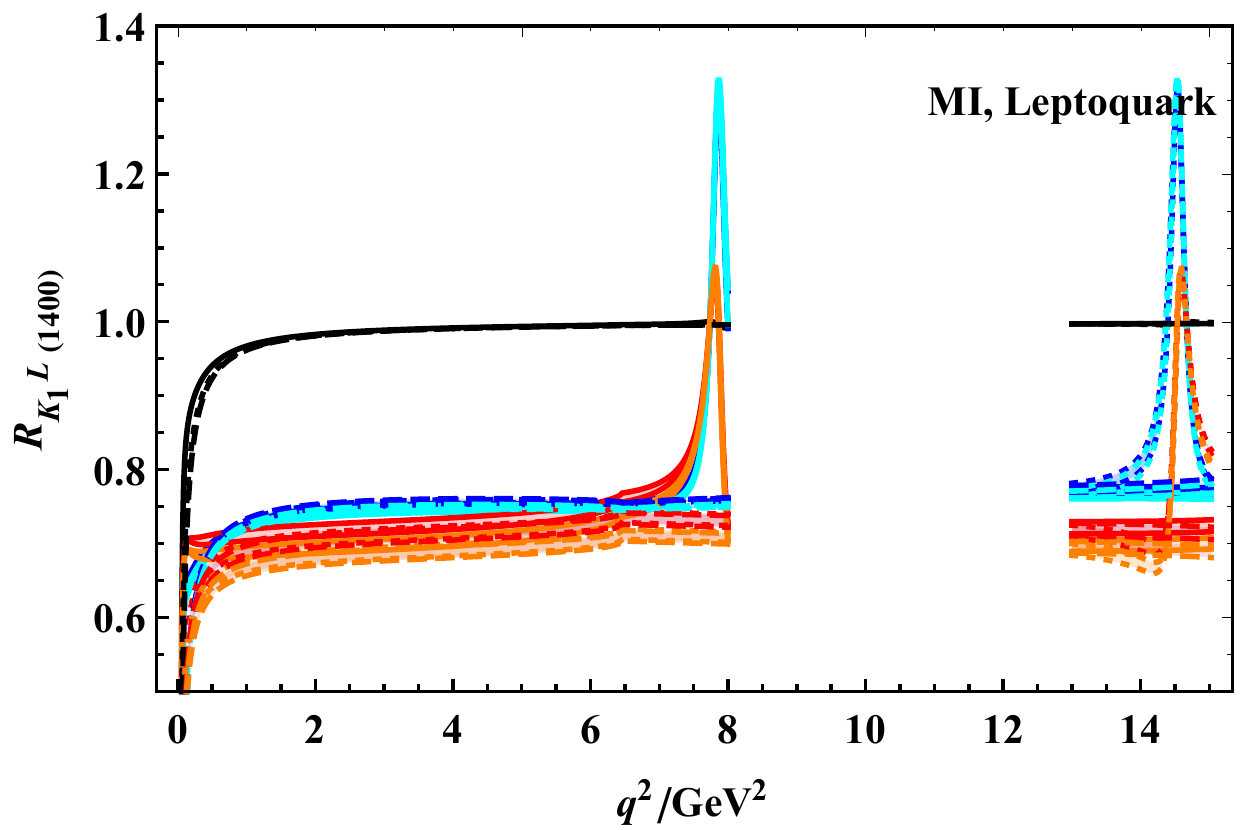}
\includegraphics[scale=0.6]{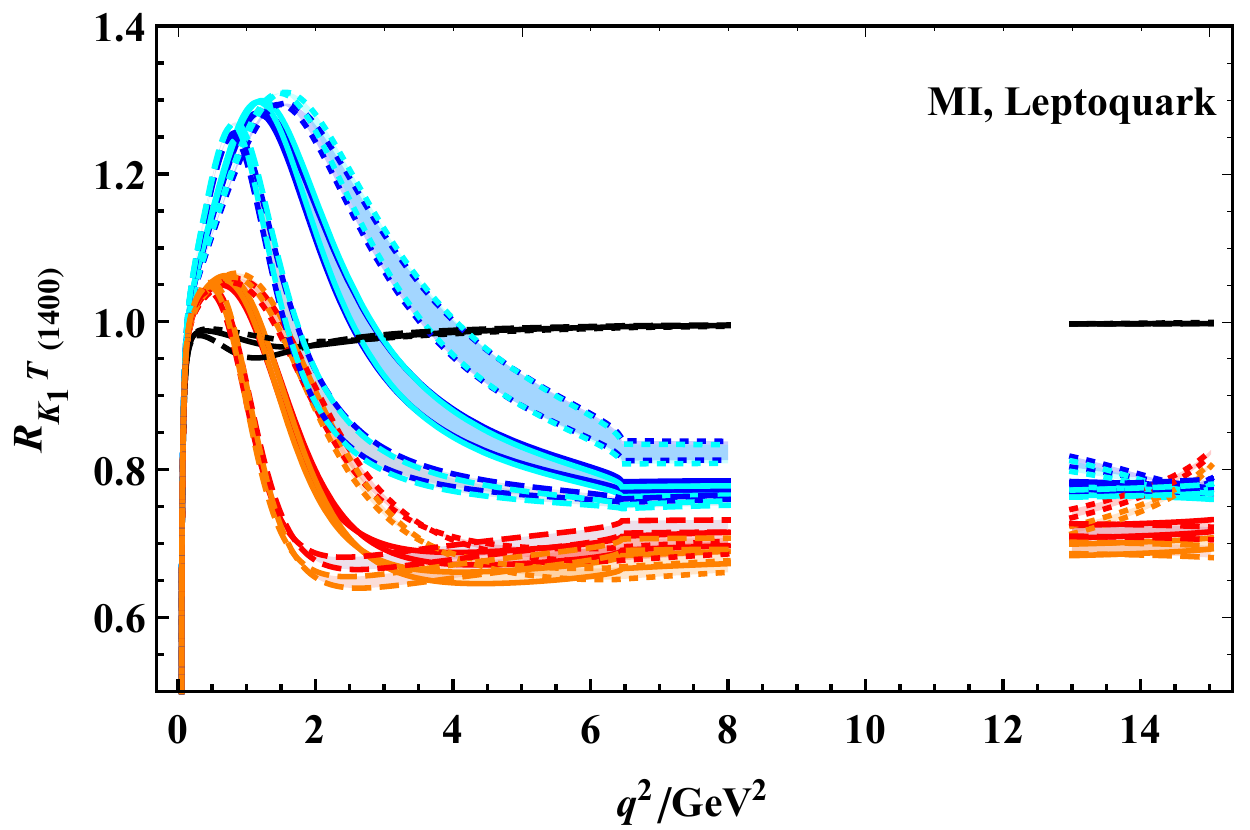}
\includegraphics[scale=0.6]{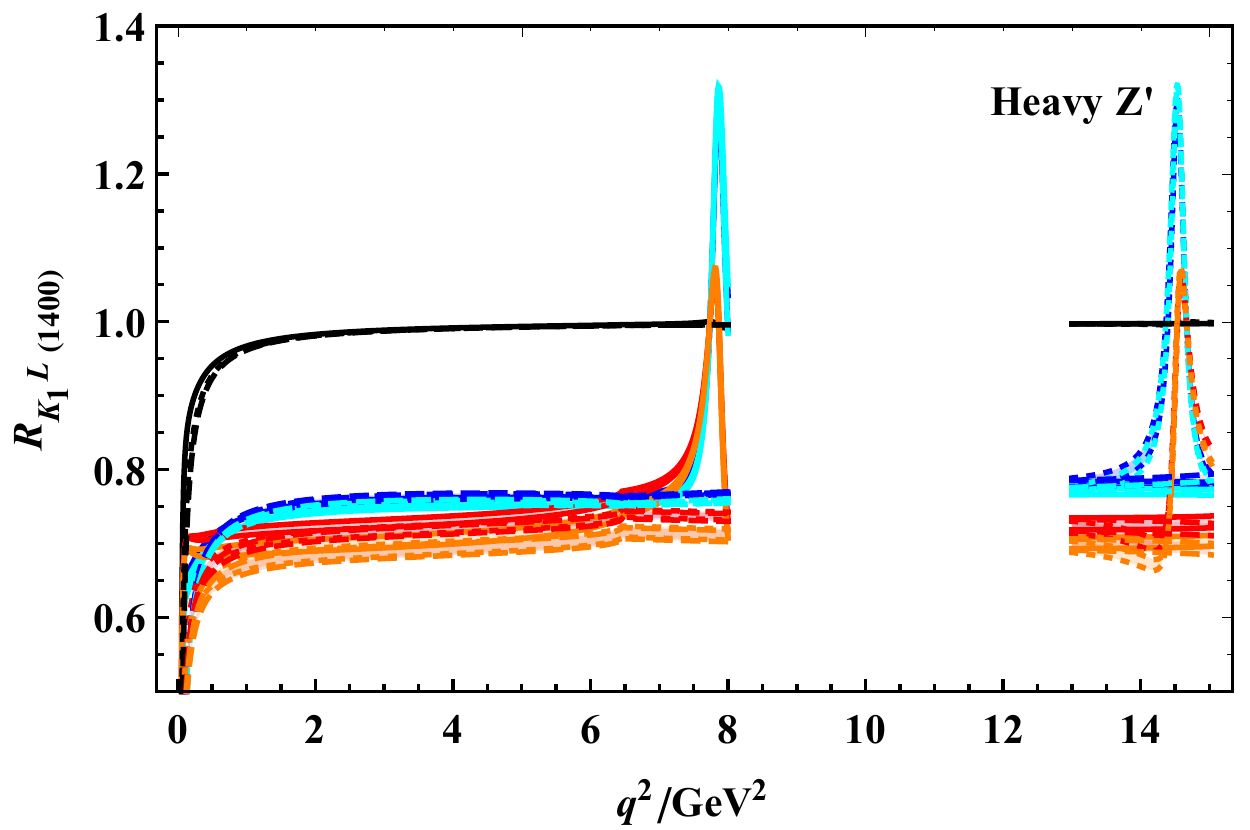}
\includegraphics[scale=0.6]{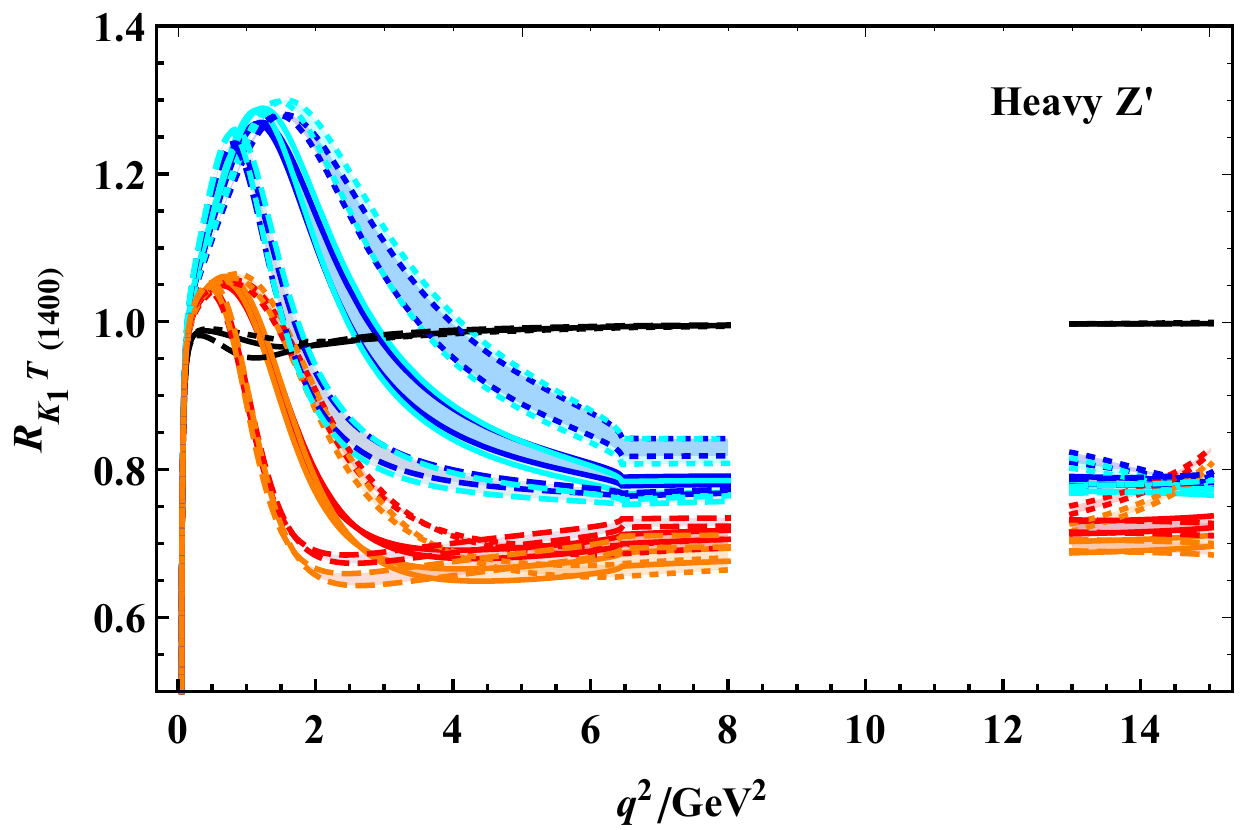}
\includegraphics[scale=0.6]{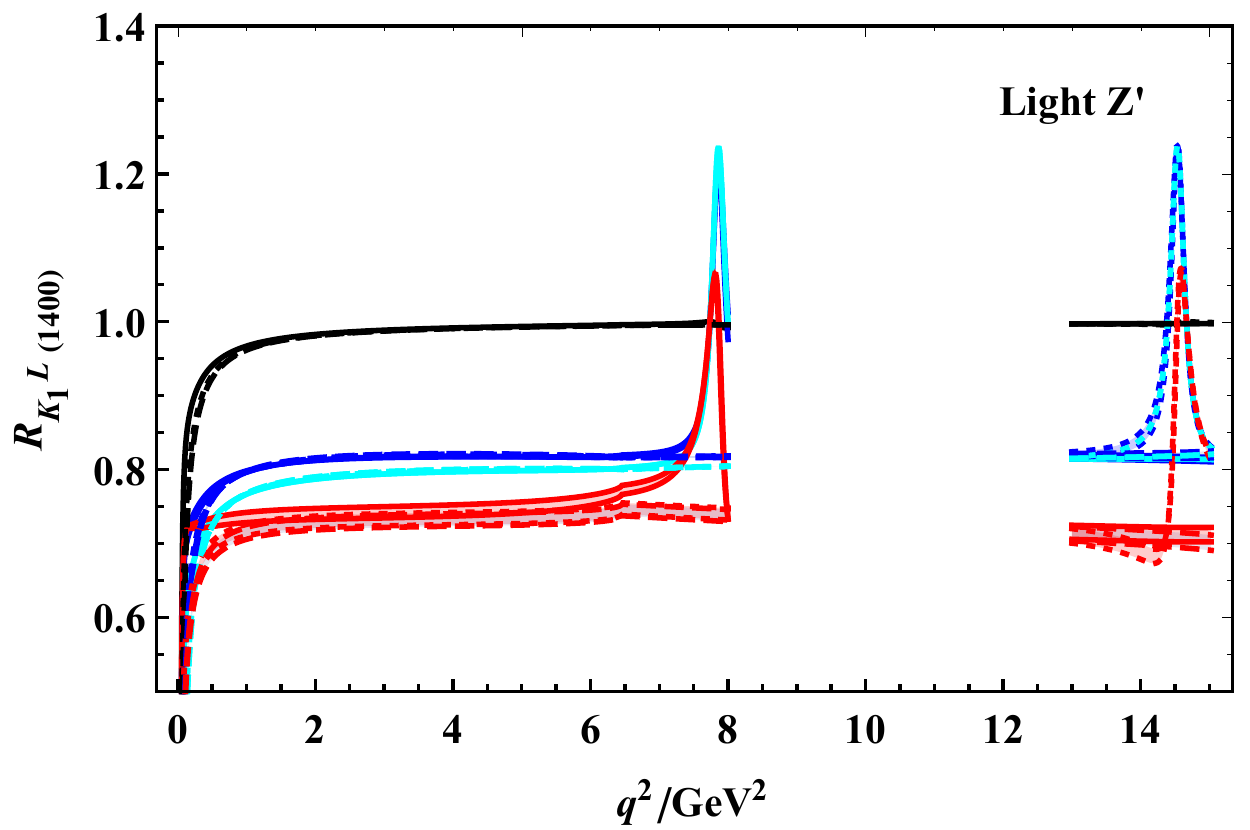}
\includegraphics[scale=0.6]{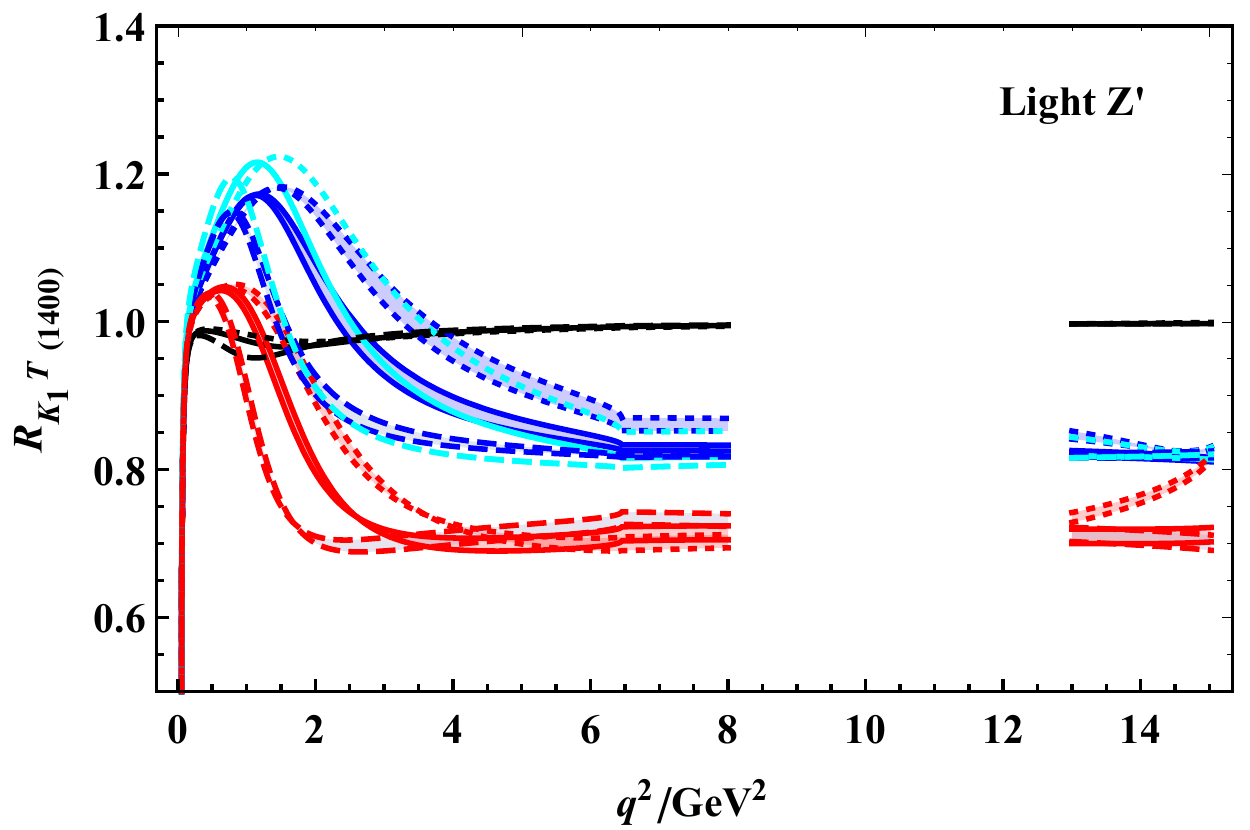}
\caption{Predictions for the polarized differential lepton universality ratio $R_{K^{L,T}_1(1400)}$ in the Standard Model, Model-indepenent, leptoquark, heavy and light $Z^\prime$ models. The legends are same as in Fig.~\ref{fig:RK11270}}
\label{fig:RK1p1400}
\end{center}
\end{figure}

For the sake of completeness and complementarity, we would also like to see the influence of NP on the values of the LFU parameters when the final state axial vector meson is $K_{1}(1400)$, which is an axial partner of the $K_{1}(1270)$.
Before presenting the results for polarized and unpolarized LFU parameters $R_{K_{1}^{(L,T)}}(1400)$, we need to recall
that $\mathcal{B}(B\to K_1(1400)\mu^+\mu^-)$ is 1-2 orders of magnitude suppressed
as compared with $\mathcal{B}(B\to K_1(1270)\mu^+\mu^-)$ ($\sim10^{-7}$). This suppression arises due to
the transformation of the transition form factors for $B\to K_{1}(1400)\mu^{+}\mu^{-}$ decay, differently than
the $B\to K_{1}(1270)\mu^{+}\mu^{-}$ decay, and was already shown in  Eqs.(\ref{K1(1270)}), (\ref{K1(1400)})
and references\cite{Hatanaka:2008xj,Hatanaka:2008gu}.
However our results in Fig.~\ref{fig:RK11400} and \ref{fig:RK1p1400} for unpolarized and polarized LFU parameters $R_{K^{(L,T)}_1(1400)}$ show even more interesting behaviours. In Fig.~\ref{fig:RK11400}, one can see that when $\theta_{K_1}=-34^\circ$ $R_{K_1(1400)}$ shows more variance and can exceed one (for scenario I) in low $q^2$ region\footnote{Note that $R_{K_1(1400)}$ is very tiny near the maximum hadronic recoil point
as shown in Fig.~\ref{fig:RK11400}, which result in the binned values less than 1 as listed in Table~\ref{tab:RK11400bin}.}.
However for $\theta_{K_1}=-45^\circ$ and $-57^\circ$ the $R_{K_1(1400)}$ does not show much variation as depicted by bands with dotted and dashed boundary lines in Fig.~\ref{fig:RK11400}. The behaviors
of $R_{K_1(1400)}$ are quite distinctive for scenario-I and II corresponding to the NP models under consideration.

For the polarized LFU parameters $R_{K^{L,T}_1(1400)}$, the results are depicted in Fig.~\ref{fig:RK1p1400}. It shown that $R_{K^{L}_1(1400)}$ is very sensitive to NP, and
very interestingly, we have found two peaks in its value, nearly at 8 GeV$^2$ slightly below the $J/\psi$ resonance and at 14.5 GeV$^2$.
These peaks arises due to the transformation of transition form factors for $B\to K_{1}(1400)$.
The peak
around $q^2=7~\rm{GeV}^2$ comes when we set the value of mixing angle $-34^\circ$ as denoted by the
solid lines and this peak shifts ahead to around $q^2=14.5~\rm{GeV}^2$  when $\theta_{K_1}=-45^\circ$ as
denoted by the dotted lines. At the value of $\theta_{K_1}=-57^\circ$ this peak goes further away.
Therefore our analysis shows that for the observable $R_{K^{L}_1(1400)}$, the position of this peak strongly depends on
the value of the mixing angle $\theta_{K_1}$. Therefore, the measurement of the value of $R_{K^{L}_1(1400)}$ can be used
to study the mixing angle. Similar to $R_{K^{T}_1(1270)}$ the value of $R_{K^{T}_1(1400)}$ is also sensitive to NP, however, this observable is more sensitive to the mixing angle $\theta_{K_1}$ as shown in Fig.~\ref{fig:RK1p1400}.

\begin{figure}[!htbp]
\begin{center}
\includegraphics[scale=0.6]{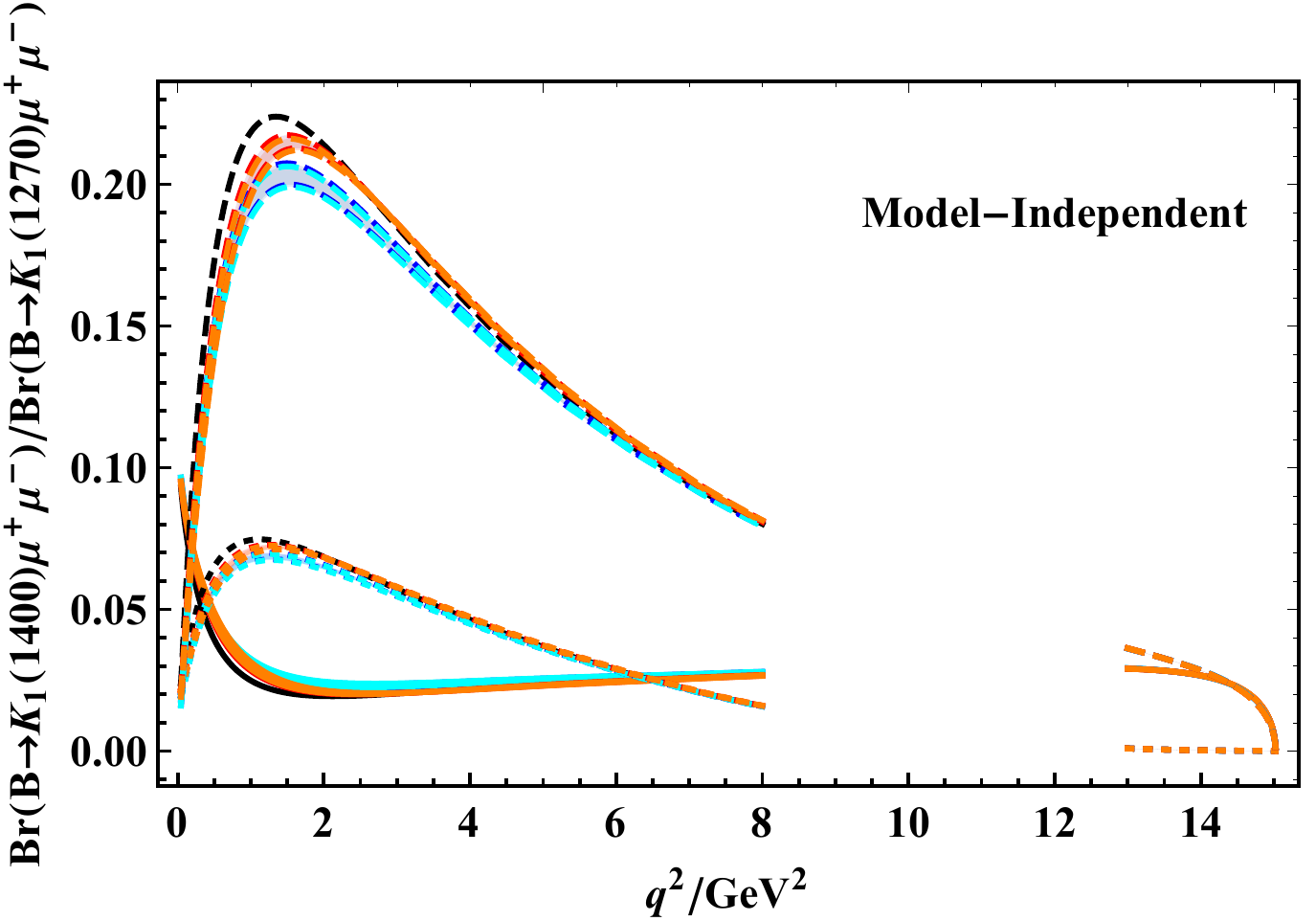}
\includegraphics[scale=0.6]{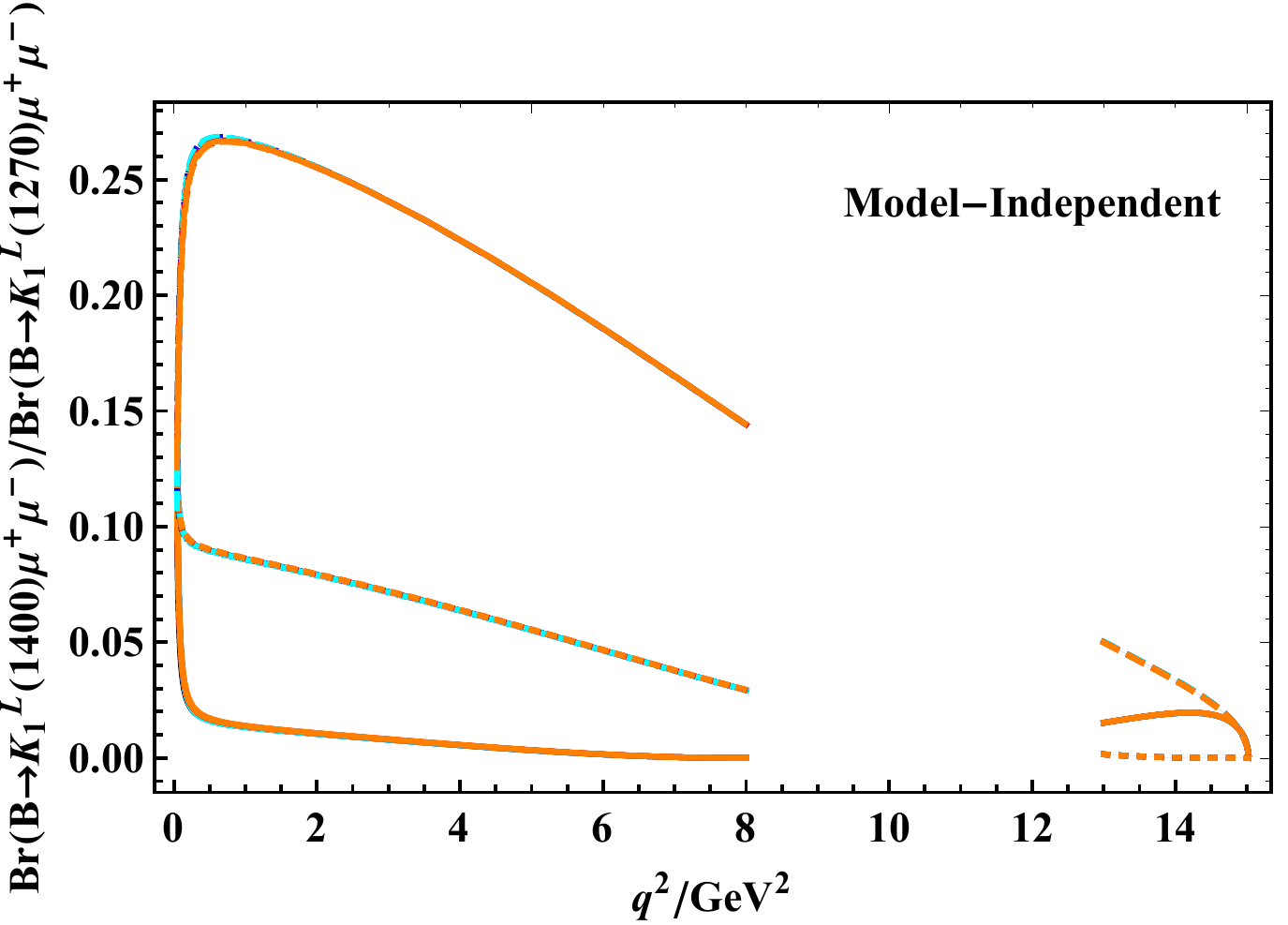}
\includegraphics[scale=0.6]{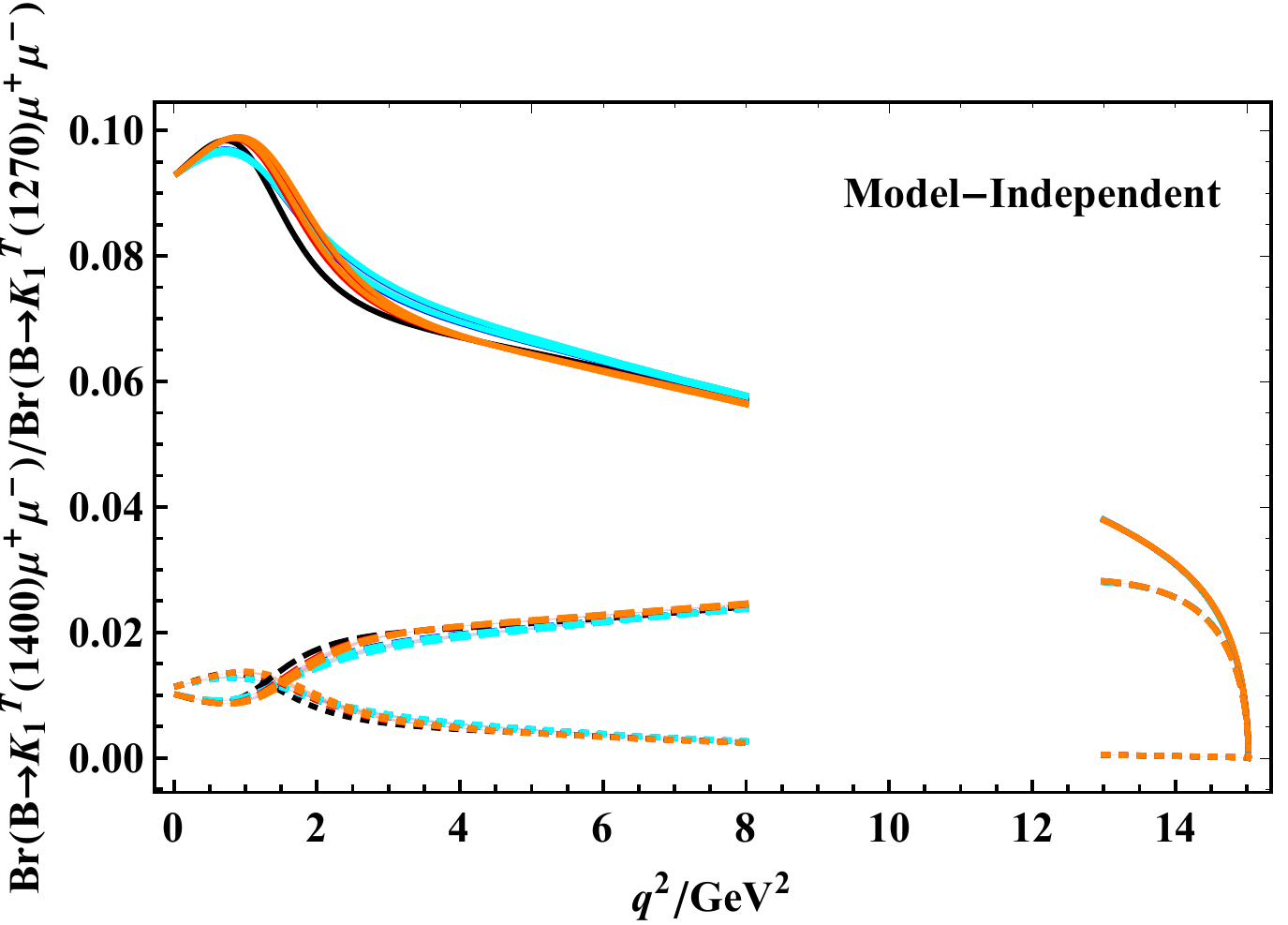}
\caption{Model-indepenent predictions for the ratio $R_\mu(K_1)=\mathcal B(B\to K_1(1400)\mu^+\mu^-)/\mathcal B(B\to K_1(1270)\mu^+\mu^-)$ for unpolarized and polarized $K_1$ in the model-independent scenarios. The legends are the same as in Fig.~\ref{fig:RK11270}}
\label{fig:RK114001270}
\end{center}
\end{figure}

As can be seen in Fig.~\ref{fig:RK11400} and \ref{fig:RK1p1400}, the LFU parameters for $K_1(1400)$ are
sensitive to $\theta_{K_1}$ in the NP scenarios/models under consideration.
To better study the NP effects, one needs other observables that can determine the $K_1$ mixing angle. As mentioned earlier,
the ratio $R_\mu(K_1)$ is possible to be such an observable, since it has already been shown to
be insensitive to the NP effects
from a single NP operator \cite{Hatanaka:2008gu}. This character also hold for more complicated NP scenarios as well as
the leptoquark and the $Z^\prime$ models. In Fig.~\ref{fig:RK114001270}, we present our results for the unpolarized and
the polarized $R_\mu$ in the model-independent scenario I and II (those for leptoquark and $Z^\prime$ models are quite similar as we have explained). These ratios are again insensitive to the NP effects: the curves of the same type with different colors almost overlap with each other. The unpolarized, longitudinal and transverse ratios $R_\mu$ can be used to determine $\theta_{K_1}$ and thus are complementary to the LFU ratios in testing the NP effects from the leptoquark and the $Z^\prime$ models.
\section{Summary and Conclusions}\label{sec6}
 Motivated by the experimental hints of the lepton universality flavor violation in the flavor-changing neutral current B decays, namely the $R_{K^{(*)}}$ anomalies, we calculate the values of unpolarized and polarized lepton flavor universality ratios $R_{K_{1}(1270,1400)}$ and $R_{K_{1}^{L,T}(1270,1400)}$ in the range of low and high $q^2$. Due to the cancellation of the hadronic uncertainties, these observables are suitable for investigating the NP effects.

 In our study, by assuming that the NP only have effects in the $b\to s\mu^+\mu^-$ transition but does not in the $b\to s e^+e^-$ transition,
 we consider different extensions of the SM, including the model-independent scenario I and II  required by
 the current $b\to s\mu^+\mu^-$ measurements, leptoquark models and heavy and light $Z^\prime$ models which
 can also satisfy scenario I and II. We use the recent constraints on the parametric values of the models under
 consideration to study how the values of the observables, mentioned above, change under the influence of NP.
 These observables against the square of the momentum transfer, $q^2$, are drawn in Fig.~\ref{fig:RK11270}-\ref{fig:RK1p1400}.

 Our study shows that this analysis on one side is the complementary check of the $R_{K^{(*)}}$ anomalies in that such kind
 of anomalies could also be seen in $R_{K_1}$. On the other hand the observables
 $R_{K_{1}(1270,1400)}$ and $R_{K_{1}^{L,T}(1270,1400)}$ are found to be more interesting and sophisticated
 for the NP due to the involvement of the mixing angle $\theta_{K_1}$. This analysis shows that in the NP scenarios and
 NP models under consideration, the results of $R_{K_1(1270)}$ are quite similar to $R_{K^{*}}$ in the sense
 that they are lower than 1 in low $q^2$ region. This feature also hold for the longitudinal $R_{K^L(1270)}$,
 while in the same region the transverse $R_{K^T(1270)}$ are greater than 1, and particularly the ratios in scenario I
 (which can only be realized in $Z^\prime$ models) can reach 1.2 or even higher.
 All the unpolarized and polarized ratios for $K_1(1270)$ are shown to be insensitive to $K_1$ mixing angle $\theta_{K_1}$
 and their values in the SM and in different NP scenarios (models) are distinguishable.

 In addition, the results of $R_{K_{1}(1400)}$ and $R_{K_{1}^{L,T}(1400)}$ are more involved because
 these ratios are sensitive to not only the NP effects but also the $K_1$ mixing angle. Therefore to better study the
 NP effects via the $R_{K_{1}^{(L,T)}(1400)}$, one essentially needs more precise value of $\theta_{K_1}$. The most notable
 characteristic of LFU parameter for $K_1(1400)$ is probably that the $R_{K^L(1270)}$ can present a peak in medium $q^2$
 region (below the resonance region) or high $q^2$ region, depending on the value of $\theta_{K_1}$.
 As a complementary study of the NP, we also perform a study of the ratio $R_\mu(K_1)$, which are found
 to be insensitive to the NP effects from the NP scenarios (models) under consideration.
 This ratio can also be used to extract the precise value of the mixing angle $\theta_{K_1}$. Therefore, if measurable,
 $R_\mu$ and $R_{K_1(1400)}$ can be complementary observables to determine the $K_1$ mixing angle and to test the leptoquark
 and $Z^\prime$ models.

 In summary, the observables considered in the current study is not only important for the complementary check on the recently found $R_{K^{(*)}}$ anomalies but also useful to extract the information of the inherent mixing angle $\theta_{K_1}$. Hence the precise measurements of $R_{K_{1}(1270,1400)}$ and $R_{K_{1}^{L,T}(1270,1400)}$ as well as $R_\mu$ in the current and the future colliders will be important for providing with insights of LFUV and as well to examine the leptoquark and $Z^\prime$ explanations of the $b\to s\mu^+\mu^-$ data.

\section{Acknowledgments}
The authors would like to thanks Bhubanjyoti Bhattacharya for useful discussions on the NP parameters in the light $Z^\prime$ models. The Authors would also like to Thanks
Faisal Munir Bhutta for providing a source file for Wilson coefficents $C_{7}^{eff}$ and $C_{9}^{eff}$. Two of the authors I. Ahmed and A. Paracha thank the hospitality provided by IHEP where most of the work is done. This work is partly supported by the National Natural Science Foundation of China (NSFC) with Grant No. 11521505 and 11621131001 and the China Postdoctoral Science Foundation with Grant No. 2018M631572.

\begin{appendix}
\section{Predictions for $R_{K_1}$ in Different Bins}
\label{appA}
In this appendix, we present our predictions for $R_{K_1(1270,1400)}$ in the SM, the model-independent scenarios and the leptoquark and the Z' models. We show the results in different $q^2$ bins in Table~\ref{tab:RK11270bin} and \ref{tab:RK11400bin}.
\begin{table*}[!htbp]
	\begin{center}
        \caption{SM and NP Predictions for the LFU ratios $R_{K_1(1270)}$ in different bins. The errors are due to the errors of the best-fit Wilson coefficients.}
			\begin{tabular}{ccccc}
				\hline
				Scenario & $\theta_{K_1}$ & $q^2$/GeV$^2: [0.045,1]$ & $q^2$/GeV$^2$: [1,6]  & $q^2$/GeV$^2$: [14,max] \\
				\hline
				SM & $-34^\circ$ & 0.881 & 0.986 & 0.997 \\
                SM & $-45^\circ$ & 0.882 & 0.986 & 0.997 \\
                SM & $-57^\circ$ & 0.883 & 0.986 & 0.997 \\		
                MI,I(A) & $-34^\circ$ & $0.782^{+0.004}_{-0.005}$ & $0.780^{+0.006}_{-0.007}$ & $0.775^{+0.004}_{-0.005}$ \\
                MI,I(A) & $-45^\circ$ & $0.796^{+0.004}_{-0.005}$ & $0.783^{+0.006}_{-0.007}$ & $0.775^{+0.004}_{-0.005}$ \\
                MI,I(A) & $-57^\circ$ & $0.811^{+0.005}_{-0.005}$ & $0.788^{+0.007}_{-0.008}$ & $0.775^{+0.004}_{-0.005}$ \\	
                MI,LQ,II(A) & $-34^\circ$ & $0.751^{+0.001}_{-0.001}$ & $0.708^{+0.008}_{-0.008}$ & $0.719^{+0.009}_{-0.008}$ \\
                MI,LQ,II(A) & $-45^\circ$ & $0.764^{+0.002}_{-0.002}$ & $0.708^{+0.008}_{-0.008}$ & $0.719^{+0.009}_{-0.008}$ \\
                MI,LQ,II(A) & $-57^\circ$ & $0.778^{+0.002}_{-0.002}$ & $0.708^{+0.008}_{-0.008}$ & $0.720^{+0.009}_{-0.008}$ \\
                MI,I(B) & $-34^\circ$ & $0.779^{+0.004}_{-0.005}$ & $0.773^{+0.006}_{-0.007}$ & $0.767^{+0.004}_{-0.005}$ \\
                MI,I(B) & $-45^\circ$ & $0.793^{+0.004}_{-0.005}$ & $0.777^{+0.006}_{-0.007}$ & $0.767^{+0.004}_{-0.005}$ \\
                MI,I(B) & $-57^\circ$ & $0.809^{+0.005}_{-0.005}$ & $0.782^{+0.007}_{-0.008}$ & $0.767^{+0.004}_{-0.005}$ \\	
                MI,LQ,II(B) & $-34^\circ$ & $0.740^{+0.001}_{-0.001}$ & $0.684^{+0.008}_{-0.007}$ & $0.695^{+0.008}_{-0.008}$ \\
                MI,LQ,II(B) & $-45^\circ$ & $0.753^{+0.001}_{-0.002}$ & $0.684^{+0.007}_{-0.007}$ & $0.695^{+0.008}_{-0.008}$ \\
                MI,LQ,II(B) & $-57^\circ$ & $0.769^{+0.002}_{-0.002}$ & $0.684^{+0.007}_{-0.007}$ & $0.695^{+0.008}_{-0.007}$ \\			
                TeV Z',I(A) & $-34^\circ$ & $0.785^{+0.004}_{-0.004}$ & $0.786^{+0.005}_{-0.006}$ & $0.782^{+0.005}_{-0.006}$ \\
                TeV Z',I(A) & $-45^\circ$ & $0.798^{+0.004}_{-0.004}$ & $0.790^{+0.006}_{-0.007}$ & $0.782^{+0.004}_{-0.005}$ \\
                TeV Z',I(A) & $-57^\circ$ & $0.813^{+0.004}_{-0.004}$ & $0.794^{+0.006}_{-0.007}$ & $0.782^{+0.004}_{-0.005}$ \\	
                TeV Z',II(A) & $-34^\circ$ & $0.754^{+0.001}_{-0.001}$ & $0.714^{+0.005}_{-0.005}$ & $0.724^{+0.008}_{-0.008}$ \\
                TeV Z',II(A) & $-45^\circ$ & $0.766^{+0.001}_{-0.001}$ & $0.713^{+0.005}_{-0.005}$ & $0.725^{+0.008}_{-0.008}$ \\
                TeV Z',II(A) & $-57^\circ$ & $0.780^{+0.001}_{-0.001}$ & $0.713^{+0.005}_{-0.005}$ & $0.725^{+0.008}_{-0.008}$ \\
                TeV Z',I(B) & $-34^\circ$ & $0.781^{+0.005}_{-0.006}$ & $0.778^{+0.007}_{-0.009}$ & $0.773^{+0.004}_{-0.005}$ \\
                TeV Z',I(B) & $-45^\circ$ & $0.795^{+0.006}_{-0.006}$ & $0.781^{+0.008}_{-0.010}$ & $0.773^{+0.004}_{-0.005}$ \\
                TeV Z',I(B) & $-57^\circ$ & $0.811^{+0.006}_{-0.006}$ & $0.786^{+0.009}_{-0.011}$ & $0.772^{+0.004}_{-0.005}$ \\	
                TeV Z',II(B) & $-34^\circ$ & $0.742^{+0.001}_{-0.001}$ & $0.688^{+0.008}_{-0.008}$ & $0.698^{+0.008}_{-0.007}$ \\
                TeV Z',II(B) & $-45^\circ$ & $0.755^{+0.002}_{-0.002}$ & $0.688^{+0.008}_{-0.008}$ & $0.699^{+0.008}_{-0.007}$ \\
                TeV Z',II(B) & $-57^\circ$ & $0.770^{+0.002}_{-0.002}$ & $0.688^{+0.008}_{-0.007}$ & $0.699^{+0.008}_{-0.007}$ \\			
                GeV Z',I(A) & $-34^\circ$ & $0.809^{+0.002}_{-0.002}$ & $0.833^{+0.003}_{-0.004}$ & $0.816^{+0.003}_{-0.004}$ \\
                GeV Z',I(A) & $-45^\circ$ & $0.819^{+0.002}_{-0.002}$ & $0.836^{+0.004}_{-0.004}$ & $0.816^{+0.003}_{-0.004}$ \\
                GeV Z',I(A) & $-57^\circ$ & $0.830^{+0.002}_{-0.003}$ & $0.838^{+0.004}_{-0.005}$ & $0.816^{+0.003}_{-0.004}$ \\	
                GeV Z',II(A) & $-34^\circ$ & $0.764^{+0.001}_{-0.001}$ & $0.727^{+0.008}_{-0.008}$ & $0.708^{+0.010}_{-0.010}$ \\
                GeV Z',II(A) & $-45^\circ$ & $0.775^{+0.001}_{-0.002}$ & $0.727^{+0.008}_{-0.007}$ & $0.708^{+0.010}_{-0.010}$ \\
                GeV Z',II(A) & $-57^\circ$ & $0.788^{+0.002}_{-0.002}$ & $0.727^{+0.008}_{-0.007}$ & $0.708^{+0.010}_{-0.010}$ \\
                MeV Z',I(A) & $-34^\circ$ & $0.788$ & $0.816$ & $0.816$ \\
                MeV Z',I(A) & $-45^\circ$ & $0.801$ & $0.818$ & $0.816$ \\
                MeV Z',I(A) & $-57^\circ$ & $0.815$ & $0.822$ & $0.816$ \\
                \hline
	\end{tabular}		
			\label{tab:RK11270bin}
	\end{center}
\end{table*}

\begin{table*}[htbp]
	\begin{center}
        \caption{SM and NP Predictions for the LFU ratio $R_{K_1(1400)}$ in different bins. The errors are due to the errors in the best-fit results of the Wilson coefficients.}
			\begin{tabular}{ccccc}
				\hline
  Scenario & $\theta_{K_1}$ & $q^2$/GeV$^2: [0.045,1]$ & $q^2$/GeV$^2$: [1,6]  & $q^2$/GeV$^2$: [13,max] \\		\hline	
				SM & $-34^\circ$ & 0.887 & 0.984 & 0.993 \\
                SM & $-45^\circ$ & 0.878 & 0.986 & 0.995 \\
                SM & $-57^\circ$ & 0.875 & 0.986 & 0.995 \\		
                MI,I(A) & $-34^\circ$ & $0.899^{+0.003}_{-0.003}$ & $0.938^{+0.018}_{-0.018}$ & $0.775^{+0.004}_{-0.005}$ \\
                MI,I(A) & $-45^\circ$ & $0.680^{+0.000}_{-0.000}$ & $0.759^{+0.003}_{-0.004}$ & $0.791^{+0.006}_{-0.006}$ \\
                MI,I(A) & $-57^\circ$ & $0.660^{+0.000}_{-0.001}$ & $0.754^{+0.003}_{-0.004}$ & $0.776^{+0.005}_{-0.006}$ \\	
                MI,LQ,II(A) & $-34^\circ$ & $0.867^{+0.002}_{-0.002}$ & $0.716^{+0.005}_{-0.005}$ & $0.719^{+0.009}_{-0.008}$ \\
                MI,LQ,II(A) & $-45^\circ$ & $0.667^{+0.005}_{-0.005}$ & $0.713^{+0.008}_{-0.008}$ & $0.739^{+0.007}_{-0.006}$ \\
                MI,LQ,II(A) & $-57^\circ$ & $0.636^{+0.007}_{-0.007}$ & $0.709^{+0.009}_{-0.008}$ & $0.717^{+0.009}_{-0.008}$ \\
                MI,I(B) & $-34^\circ$ & $0.900^{+0.003}_{-0.003}$ & $0.866^{+0.015}_{-0.016}$ & $0.767^{+0.004}_{-0.005}$ \\
                MI,I(B) & $-45^\circ$ & $0.672^{+0.000}_{-0.001}$ & $0.751^{+0.003}_{-0.004}$ & $0.784^{+0.005}_{-0.006}$ \\
                MI,I(B) & $-57^\circ$ & $0.653^{+0.000}_{-0.001}$ & $0.747^{+0.003}_{-0.004}$ & $0.769^{+0.005}_{-0.006}$ \\	
                MI,LQ,II(B) & $-34^\circ$ & $0.865^{+0.002}_{-0.002}$ & $0.693^{+0.004}_{-0.004}$ & $0.695^{+0.008}_{-0.007}$ \\
                MI,LQ,II(B) & $-45^\circ$ & $0.648^{+0.004}_{-0.004}$ & $0.689^{+0.008}_{-0.007}$ & $0.716^{+0.006}_{-0.006}$ \\
                MI,LQ,II(B) & $-57^\circ$ & $0.616^{+0.007}_{-0.006}$ & $0.685^{+0.008}_{-0.008}$ & $0.693^{+0.008}_{-0.008}$ \\			
                TeV Z',I(A) & $-34^\circ$ & $0.899^{+0.003}_{-0.003}$ & $0.871^{+0.013}_{-0.015}$ & $0.782^{+0.004}_{-0.005}$ \\
                TeV Z',I(A) & $-45^\circ$ & $0.686^{+0.000}_{-0.000}$ & $0.766^{+0.003}_{-0.004}$ & $0.798^{+0.004}_{-0.005}$ \\
                TeV Z',I(A) & $-57^\circ$ & $0.667^{+0.000}_{-0.001}$ & $0.762^{+0.003}_{-0.003}$ & $0.783^{+0.004}_{-0.005}$ \\	
                TeV Z',II(A) & $-34^\circ$ & $0.867^{+0.002}_{-0.002}$ & $0.721^{+0.003}_{-0.003}$ & $0.725^{+0.008}_{-0.008}$ \\
                TeV Z',II(A) & $-45^\circ$ & $0.671^{+0.003}_{-0.003}$ & $0.718^{+0.006}_{-0.005}$ & $0.744^{+0.006}_{-0.006}$ \\
                TeV Z',II(A) & $-57^\circ$ & $0.641^{+0.005}_{-0.005}$ & $0.714^{+0.006}_{-0.006}$ & $0.723^{+0.009}_{-0.008}$ \\
                TeV Z',I(B) & $-34^\circ$ & $0.900^{+0.004}_{-0.004}$ & $0.868^{+0.019}_{-0.021}$ & $0.773^{+0.004}_{-0.005}$ \\
                TeV Z',I(B) & $-45^\circ$ & $0.677^{+0.000}_{-0.001}$ & $0.757^{+0.004}_{-0.005}$ & $0.789^{+0.005}_{-0.005}$ \\
                TeV Z',I(B) & $-57^\circ$ & $0.658^{+0.000}_{-0.001}$ & $0.752^{+0.004}_{-0.005}$ & $0.774^{+0.004}_{-0.005}$ \\	
                TeV Z',II(B) & $-34^\circ$ & $0.866^{+0.002}_{-0.002}$ & $0.696^{+0.004}_{-0.005}$ & $0.699^{+0.008}_{-0.007}$ \\
                TeV Z',II(B) & $-45^\circ$ & $0.651^{+0.005}_{-0.005}$ & $0.693^{+0.008}_{-0.008}$ & $0.719^{+0.006}_{-0.006}$ \\
                TeV Z',II(B) & $-57^\circ$ & $0.619^{+0.007}_{-0.007}$ & $0.688^{+0.009}_{-0.008}$ & $0.697^{+0.008}_{-0.008}$ \\			
                GeV Z',I(A) & $-34^\circ$ & $0.893^{+0.002}_{-0.002}$ & $0.891^{+0.009}_{-0.010}$ & $0.818^{+0.003}_{-0.004}$ \\
                GeV Z',I(A) & $-45^\circ$ & $0.737^{+0.001}_{-0.002}$ & $0.820^{+0.001}_{-0.002}$ & $0.831^{+0.003}_{-0.004}$ \\
                GeV Z',I(A) & $-57^\circ$ & $0.722^{+0.002}_{-0.002}$ & $0.817^{+0.001}_{-0.002}$ & $0.819^{+0.003}_{-0.004}$ \\	
                GeV Z',II(A) & $-34^\circ$ & $0.869^{+0.002}_{-0.002}$ & $0.733^{+0.005}_{-0.005}$ & $0.711^{+0.010}_{-0.009}$ \\
                GeV Z',II(A) & $-45^\circ$ & $0.687^{+0.004}_{-0.004}$ & $0.733^{+0.008}_{-0.008}$ & $0.732^{+0.008}_{-0.007}$ \\
                GeV Z',II(A) & $-57^\circ$ & $0.660^{+0.007}_{-0.006}$ & $0.729^{+0.009}_{-0.008}$ & $0.709^{+0.010}_{-0.010}$ \\
                MeV Z',I(A) & $-34^\circ$ & $0.895$ & $0.884$ & $0.816$ \\
                MeV Z',I(A) & $-45^\circ$ & $0.690$ & $0.800$ & $0.830$ \\
                MeV Z',I(A) & $-57^\circ$ & $0.677$ & $0.796$ & $0.817$ \\				
                \hline	
			\end{tabular}		
			\label{tab:RK11400bin}
	\end{center}
\end{table*}
\end{appendix}

\end{document}